\documentclass[conference]{IEEEtran}

\usepackage{breakurl}
\usepackage{amsmath,amssymb,amsfonts}
\usepackage{mathtools}
\usepackage{algorithmic}
\usepackage{graphicx}
\usepackage{textcomp}
\usepackage[dvipsnames]{xcolor}
\usepackage{multirow}
\usepackage{todonotes}
\usepackage{natbib}
\usepackage{comment}
\usepackage{hyperref}
\usepackage{listings}
\usepackage{xcolor}
\usepackage{subcaption}
\usepackage{array}
\usepackage{microtype}

\usepackage{tikz}
\newcommand*\circled[1]{\raisebox{.5pt}{\textcircled{\raisebox{-.9pt} {#1}}}}

\lstdefinestyle{cpp}{
  language=C++,
  basicstyle=\ttfamily\small, 
  morekeywords={Func,RDom, Var,ImageParam},
  keywordstyle=\color{blue}\bfseries,
  stringstyle=\color{orange},
  commentstyle=\color{gray}\itshape,
  numbers=left,
  numberstyle=\tiny\color{gray},
  stepnumber=1,
  numbersep=10pt,
  backgroundcolor=\color{white},
  showspaces=false,
  showstringspaces=false,
  showtabs=false,
  frame=single,
  tabsize=2,
  captionpos=b,
  breaklines=true,
  breakatwhitespace=false,
  escapeinside={(*@}{@*)},
  xleftmargin=2em,frame=single,framexleftmargin=1.5em,
  literate=
    *{x512}{{{x512}}}{4}
     {x256}{{{x256}}}{4}
     {x16}{{{x16}}}{3}
     {x18}{{{x18}}}{3}
     {x32}{{{x32}}}{3}
     {x4}{{{x4}}}{2}
     {x8}{{{x8}}}{2}
     {x3}{{{x3}}}{2}
     {x6}{{{x6}}}{2}
     {float32x8}{{{float32x8}}}{9}
     {float32x512}{{{float32x512}}}{11}
     {float32x256}{{{float32x256}}}{11}
     {float32x8192}{{{float32x8192}}}{12}
     {float32x2048}{{{float32x2048}}}{12}
     {float16}{{{float16}}}{7}
     {m32n8k16.f16}{{{m32n8k16.f16}}}{12}
     {m32n8k16.f32.f32}{{{m32n8k16.f32.f32}}}{16}
     {0}{{{\color{ForestGreen}0}}}{1}
     {1}{{{\color{ForestGreen}1}}}{1}
     {2}{{{\color{ForestGreen}2}}}{1}
     {3}{{{\color{ForestGreen}3}}}{1}
     {4}{{{\color{ForestGreen}4}}}{1}
     {5}{{{\color{ForestGreen}5}}}{1}
     {6}{{{\color{ForestGreen}6}}}{1}
     {7}{{{\color{ForestGreen}7}}}{1}
     {8}{{{\color{ForestGreen}8}}}{1}
     {9}{{{\color{ForestGreen}9}}}{1}
     {.0}{{{\color{ForestGreen}.0}}}{2}
     {.1}{{{\color{ForestGreen}.1}}}{2}
     {.2}{{{\color{ForestGreen}.2}}}{2}
     {.3}{{{\color{ForestGreen}.3}}}{2}
     {.4}{{{\color{ForestGreen}.4}}}{2}
     {.5}{{{\color{ForestGreen}.5}}}{2}
     {.6}{{{\color{ForestGreen}.6}}}{2}
     {.7}{{{\color{ForestGreen}.7}}}{2}
     {.8}{{{\color{ForestGreen}.8}}}{2}
     {.9}{{{\color{ForestGreen}.9}}}{2}
     {.f}{{{\color{ForestGreen}.f}}}{2}
}
\lstdefinestyle{python}{
  language=Python,
  basicstyle=\ttfamily\small, 
  morekeywords={Func,RDom, Var,ImageParam},
  keywordstyle=\color{blue}\bfseries,
  stringstyle=\color{orange},
  commentstyle=\color{gray},
  numbers=left,
  numberstyle=\tiny\color{gray},
  stepnumber=1,
  numbersep=10pt,
  backgroundcolor=\color{white},
  showspaces=false,
  showstringspaces=false,
  showtabs=false,
  frame=single,
  tabsize=2,
  captionpos=b,
  breaklines=true,
  breakatwhitespace=false,
  escapeinside={(*@}{@*)},
  xleftmargin=2em,frame=single,framexleftmargin=1.5em
}
\definecolor{DarkPurple}{rgb}{0.4, 0.0, 0.4}
\definecolor{ForestGreen}{RGB}{34,139,34}
\lstdefinelanguage{egglog}{
  sensitive,
  basicstyle=\ttfamily\small, 
  commentstyle=\color{gray}\itshape,
  morecomment=[l]{;},
  moredelim=[s][{\itshape\color[rgb]{0,0,0.75}}]{\#[}{]},
  morestring=[b]{"},
  alsodigit={0123456789},
  alsoother={},
  alsoletter={!},
  alsoletter={:},
  alsoletter={-},
  alsoletter=0123456789,
  otherkeywords={=>},
  otherkeywords={:-},
  otherkeywords={|},
  otherkeywords={.},
  morekeywords={assert, function, relation, datatype, rewrite,
                rule, union, merge, default, when,
                check, set, run, sort, define, extract, let},
  keywordstyle=\color{blue}\bfseries,
  otherkeywords={:merge, :default, :when},
  morestring=[b]",
  stringstyle=\color{ForestGreen},
  moredelim=[is][\color{DarkPurple}\texttt]{[}{]},
belowskip=0.2em,
aboveskip=0.2em,
  literate=
    *{AMX2Mem}{{{AMX2Mem}}}{7}
     {Mem2AMX}{{{Mem2AMX}}}{7}
     {i64}{{{i64}}}{3}
     {f64}{{{f64}}}{3}
     {Float32}{{{Float32}}}{7}
     {Float16}{{{Float16}}}{7}
     {BFloat16}{{{BFloat16}}}{8}
     {t1}{{{t1}}}{2}
     {t2}{{{t2}}}{2}
     {l1}{{{l1}}}{2}
     {l2}{{{l2}}}{2}
     {e1}{{{e1}}}{2}
     {e2}{{{e2}}}{2}
     {0}{{{\color{ForestGreen}0}}}{1}
     {1}{{{\color{ForestGreen}1}}}{1}
     {2}{{{\color{ForestGreen}2}}}{1}
     {3}{{{\color{ForestGreen}3}}}{1}
     {4}{{{\color{ForestGreen}4}}}{1}
     {5}{{{\color{ForestGreen}5}}}{1}
     {6}{{{\color{ForestGreen}6}}}{1}
     {7}{{{\color{ForestGreen}7}}}{1}
     {8}{{{\color{ForestGreen}8}}}{1}
     {9}{{{\color{ForestGreen}9}}}{1}
     {.0}{{{\color{ForestGreen}.0}}}{2}
     {.1}{{{\color{ForestGreen}.1}}}{2}
     {.2}{{{\color{ForestGreen}.2}}}{2}
     {.3}{{{\color{ForestGreen}.3}}}{2}
     {.4}{{{\color{ForestGreen}.4}}}{2}
     {.5}{{{\color{ForestGreen}.5}}}{2}
     {.6}{{{\color{ForestGreen}.6}}}{2}
     {.7}{{{\color{ForestGreen}.7}}}{2}
     {.8}{{{\color{ForestGreen}.8}}}{2}
     {.9}{{{\color{ForestGreen}.9}}}{2}
     {.f}{{{\color{ForestGreen}.f}}}{2}
}%

\newcommand{\toolname}{\mbox{\textsc{HardBoiled}}}

\newcommand{\usl}{user-schedulable language}
\newcommand{\Usl}{User-schedulable language}

\newenvironment{revision}
    {}
    {}

\newenvironment{shepherd}
    {}
    {}

\begin{document}

\title{
Pushing Tensor Accelerators beyond MatMul in a User-Schedulable Language}


\author{\IEEEauthorblockN{Yihong Zhang}
\IEEEauthorblockA{\textit{University of Washington, USA}}\\
\and
\IEEEauthorblockN{Derek Gerstmann}
\IEEEauthorblockA{\textit{Adobe, USA}}\\
\and
\IEEEauthorblockN{Andrew Adams}
\IEEEauthorblockA{\textit{Adobe, USA}}\\
\and
\IEEEauthorblockN{Maaz Bin Safeer Ahmad}
\IEEEauthorblockA{\textit{Adobe, USA}}\\

}

\maketitle

\begin{abstract}

\begin{revision}
Tensor accelerators now represent a growing share of compute resources in modern CPUs and GPUs. However, they are hard to program, leading developers to use vendor-provided kernel libraries that support tensor accelerators. As a result, the usage of tensor accelerators is limited to the provided interface, mainly designed for traditional ML and scientific computing workloads.

In this paper, we show that tensor accelerators can improve the performance of applications beyond simple variants of MatMul. For example, many image processing pipelines are linear transformations over matrices in disguise and can therefore utilize such specialized hardware. This is nonetheless hindered by the difficulties in programming tensor accelerators. We tackle this problem with compiler-based techniques. We use the Halide \usl~and express operations as Halide algorithms succinctly. To this end, we implement a flexible tensor instruction selector based on equality saturation. The tensor instruction selector supports both CPU- and GPU-attached tensor accelerators and works with existing scheduling operations (e.g., producer-consumer fusion). Together, this enables developers to write diverse accelerator-leveraging applications in a few dozen lines.

Using our system, we demonstrate the potential of tensor accelerators beyond their traditional domains. We implement several image processing pipelines (e.g., filtering, resampling, and denoising) in our system and evaluate them against non-accelerator-leveraging baselines. We show that these pipelines can achieve significant speedups. For example, a downsampling routine is sped up by $6.1\times$  by utilizing Tensor Cores on an Nvidia RTX 4070 GPU.
\end{revision}

\end{abstract}


\section{Introduction}

\begin{revision}
The unrelenting demand for performance in AI and scientific computing has driven hardware vendors to develop highly efficient domain-specific accelerators for matrix operations, with considerable success. In fact, the ability to leverage such hardware accelerators has become essential to extracting peak performance. For example, the Tensor Cores in Nvidia’s RTX 4090 deliver roughly 2$\times$ throughput of its general-purpose CUDA cores, while datacenter GPUs like the H100 offer 8-16$\times$ more potential. Yet such architectures are highly specialized, constantly evolving, and hard to program. Instead of programming against specialized hardware, developers use vendor-provided kernel libraries that consist of hardware-specific routines.

These kernel libraries support common math operations like matrix multiplication (MatMul), but the tensor accelerators could be useful for many other tasks. For example, many operations on signals are linear transformations over inputs and can be modeled as MatMul. 
However, these tasks remain inaccessible due to the narrow, rigid interfaces of kernel libraries.
As a result, existing state-of-the-art implementations for these operations fail to utilize specialized hardware at all.
Even in cases where application developers can contort their operations and data layouts around the provided API, the rigid interface also limits their ability to perform program optimizations such as code fusion, data layout reorganization, and workload scheduling, leading to inferior implementations.

In this paper, we study performance-critical applications where tensor accelerators are underused. 
We argue \usl{}s \citep{halide, tvm} like Halide present a promising approach to expand the reach of hardware accelerators beyond the capability of kernel libraries.
Halide is a domain-specific language for high-performance applications based on the idea of decoupling algorithms from schedules. 
Developers can succinctly describe their applications in a few lines of Halide algorithms, and use schedules for fine-grained controls over machine-specific execution strategies (e.g., loop tiling, vectorization).
Ideally, developers would develop their tensor-accelerator-empowered kernels by focusing on the high-level algorithms and schedules, without worrying too much about the mapping to specific accelerator calls.


Current \usl{}~implementations lack flexible support for tensor accelerators.
State-of-the-art implementations hard-code supported syntactic patterns and use pattern-based rewriting rules.
This approach is brittle and inflexible and takes excessive developer cycles.
For example, it took Halide developers seven months to add initial support for Intel's Advanced Matrix Extensions (AMX) accelerator \citep{halide_pr_5818}. However, despite the considerable effort, it only recognizes a very limited syntactic form. A later pull request that generalized this support \citep{halide_pr_6581} took another six months to merge. Yet we found the existing implementation contained a number of bugs.

We revamp Halide's support for tensor accelerators with a new instruction selector, dubbed \toolname{}.
The goal of \toolname{} is to support flexible use of tensor accelerators for domain applications.
To this end, we use equality saturation (EqSat), a technique for rewrite-based program optimization that is robust to syntactic variations in the source program~\citep{peggy}.
We repurpose Halide's existing vector abstraction for representing tensor programs and express tensor identities as rewrite rules in equality saturation.
This provides a consistent user interface, as from the user's perspective, tensor instructions are similar to vectorized instructions---just with more than one dimension.

Using our system, we explore the extent to which applications in signal and image processing can be sped up by tensor accelerators.
As a case study, we consider the following applications: 1D/2D convolution, integer-factor resampling, non-integer-factor resampling, recursive filtering, and denoising using discrete cosine transform.
We tuned and evaluated these applications on Tensor Cores and show that they can achieve up to 6.1$\times$ speedup on convolution-like kernels, and 1.1$\times$ to 1.4$\times$ speedup on end-to-end applications.
As an additional validation, we have also implemented several kernels representative of classical ML workloads in our system and show that they are comparable to library implementations.

To summarize, we make two contributions in this work. 
First, we implement \toolname{}, a tensor instruction selector for Halide. \toolname{} can identify and compile diverse workloads for tensor accelerators and works seamlessly with Halide schedules. It supports both Intel AMX and Nvidia Tensor Cores.
Second, using our system, we implement a set of classical image processing workloads in ways that effectively utilize tensor accelerators. We show that they achieve significant speedups compared to non-accelerator-leveraging baselines under appropriate settings.
\end{revision}

\section{Background}

\subsection{Tensor Accelerators}

Modern hardware has increasingly added specialized AI accelerators to offload math-heavy tensor computations like MatMul.
For example, Nvidia's Tensor Cores are dedicated accelerators that support fast matrix operations for specific sizes and layouts.
Similarly, Intel's Advanced Matrix Extensions (AMX) support matrix operations between tile registers with sizes up to $16\times 32$.
The ability to leverage these accelerators can lead to significant speedups.

\begin{revision}
On the other hand, tensor accelerators are underutilized. Their uses are limited primarily to traditional ML/AI and scientific computing workloads. As we will show in this paper, many operations over signals are linear tensor operations and can leverage these tensor accelerators. Nonetheless, state-of-the-art kernel implementations of these operations often fail to use such specialized hardware. This is unfortunate, given the massive industrial investment in improving their throughput.

This underutilization can be partially attributed to the difficulty in programming hardware accelerators, which forces developers to use vendor's implementations of a few common high-level operations. Because library interfaces are often coarse-grained, they prevent optimizations such as producer–consumer fusion, making them ill-suited for many domain-specific applications. A more effective approach is to provide developers with a user-friendly interface for programming tensor accelerators.
\end{revision}


\subsection{User-Schedulable Languages}

In a \usl{} like Halide, developers express the desired computations as functional definitions of arrays, without regard to how those computations are executed.
At this stage, the focus is purely on the high-level computation.
For example, MatMul may be expressed in Halide as follows:
\begin{lstlisting}[style=cpp, basicstyle=\ttfamily\footnotesize]
mm(y, x) = 0.f;
mm(y, x) += A(r, x) * B(y, r);
\end{lstlisting}
In a separate stage, developers write \emph{schedules} that describe the optimization details for the computation, such as tiling, loop orders, and vectorization.

\begin{revision}
For example, the following schedule breaks down the computation of \verb|mm| into $8\times 2$ tiles. Within each tile, the $y$ axis is vectorized while the $x$ axis is unrolled.
\end{revision}

\begin{lstlisting}[style=cpp, basicstyle=\ttfamily\footnotesize]
Var xi, yi;
mm.in().tile(x, y, xi, yi, 2, 8)
  .vectorize(yi).unroll(xi);
mm.compute_at(mm.in(), y);
mm.vectorize(yi).unroll(xi);
mm.update().vectorize(yi).unroll(xi);
\end{lstlisting}

\begin{revision}
With this schedule, our matrix multiplication compiles to the following low-level code.
\end{revision}

\begin{lstlisting}[style=python, basicstyle=\ttfamily\footnotesize]
for x in range(0, N, 16):
 for y in range(0, M, 2):
  float mm0[8][2] # local 8x2 buffer
  mm0[0:8, 0] = 0;  mm0[0:8, 1] = 0
  for r in range(0, K):
   mm0[0:8, 0] += A[r, x]   * B[y:y+8, r]
   mm0[0:8, 1] += A[r, x+1] * B[y:y+8, r]
  mm[y:y+8, x]   = mm0[0:8, 0]
  mm[y:y+8, x+1] = mm0[0:8, 1]
\end{lstlisting}

\subsection{Tensor Accelerators in User-Schedulable Languages}

\begin{revision}
\Usl{}s like Halide present a promising paradigm to host developments of tensor-accelerator-based applications. Halide enables advanced optimizations, such as tiling and fusion, to be expressed in just a few lines, and it generates efficient low-level implementations for various hardware platforms. This allows developers to prototype their applications for various hardware quickly.

However, languages like Halide have only limited support for hardware accelerators. To add support, compiler developers must match all possible syntactical patterns and handle different corner cases. This typically results in fragmented, imperative implementations and special-case handling.
Currently, Halide's support for Intel AMX is limited to a small set of patterns 
and contains multiple bugs, and no support for Nvidia Tensor Cores exists. 

\end{revision}


An alternative approach is taken by TVM \citep{tvm}, which provides a user-extensible pattern-matching system for accelerator intrinsics via its tensorize directive. While this places higher demands on the user, with this system, a user can explicitly tell the compiler how some portion of their algorithm maps to an accelerator intrinsic, without having to wait for official compiler support.
\begin{revision}
In practice, programs in TVM are often auto-scheduled with works like TensorIR \citep{tensorir_2022} and Ansor \citep{ansor}. We focus on scenarios where application developers write schedules manually, as is typical in Halide's production usage. We compare the two approaches further in \autoref{sec:relatedwork}.
\end{revision}

\subsection{Equality Saturation}

\begin{figure}
    \centering
    \includegraphics[width=0.6\linewidth]{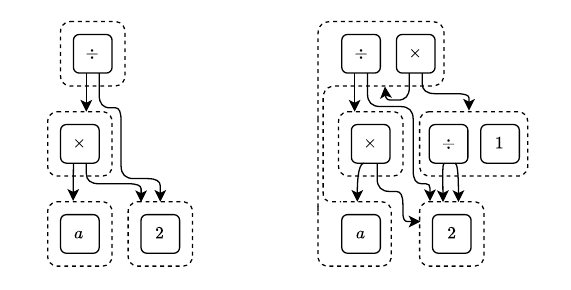}
    \vspace*{-0.2cm}
    \caption{Two example E-graphs \cite{egg}}
    \vspace*{-0.3cm}
    \label{fig:egraph-example}
\end{figure}

\begin{revision}
Our goal is to build a robust and flexible tensor instruction selector to support our exploration of accelerator-driven domain-specific applications. To this end, we use equality saturation (EqSat) \citep{egg,peggy}. \end{revision} EqSat is a program optimization technique that efficiently explores program spaces defined by rewrite rules.
It explores combinations of different rewrite orders to find the optimal program.
To mitigate the combinatorial blowup, 
it uses a data structure called E-graphs to compactly represent the equivalence classes of programs.
\autoref{fig:egraph-example} shows two example E-graphs. The E-graph on the left represents the program $(a\times 2)\div 2$, and on the right is the E-graph after rules $(a\times 2)\div 2\rightarrow a\times (2\div 2)$, $2\div 2\rightarrow 1$, and $a\times 1\rightarrow a$.
Nodes grouped by the same dashed boxes (called \emph{E-classes}) belong to the same equivalence class.
For instance, the E-graph on the right shows $(a\times 2)\div 2$ is equivalent to $a$.

In this paper, we use egglog \citep{egglog}, a recent system that combines equality saturation and deductive reasoning. Two kinds of rules are supported in egglog: \lstinline[language=egglog]{rewrite} and \lstinline[language=egglog]{rule}. For example, \lstinline[language=egglog]{(rewrite (Add [x y]) (Add [y x]))} defines commutativity of addition. A \lstinline[language=egglog]{rule} consists of a query and an action. 
For instance, the following rule defines the type derivation for function application.
\begin{lstlisting}[language=egglog, basicstyle=\ttfamily\footnotesize]
(rule ((= e (App [e1] [e2]))
       (has-type [e1] (Arrow [t1] [t2]))
       (has-type [e2] [t1]))
      ((has-type [e] [t2])))
\end{lstlisting}
In its query, this rule looks for the pattern \lstinline[language=egglog]{(App [e1 e2])} in the E-graph satisfying  {\color{DarkPurple}\verb|e1|} has type {\color{DarkPurple}\verb|t1|}$\to${\color{DarkPurple}\verb|t2|} and {\color{DarkPurple}\verb|e2|} has type {\color{DarkPurple}$\verb|t1|$}.
For each match of this pattern, the rule asserts that \lstinline[language=egglog]{(App [e1 e2])} should have type {\color{DarkPurple}\verb|t2|}. In fact, a \lstinline[language=egglog]{rewrite} is a special form of \lstinline[language=egglog]{rule} and can also be desugared using the \lstinline[language=egglog]{union} keyword, e.g.,
\lstinline[language=egglog]{(rule ((= [e ] (Add [x y]))) ((union [e ](Add [y x]))))}.

\section{Writing Tensor Kernels with HardBoiled}


In this section, we present the design of \toolname{} and the workflow of using \toolname{}-enabled Halide to write kernels that use tensor accelerators.

\toolname{} works as a compiler pass in Halide's optimization pipeline.
To compile statements to take advantage of tensor accelerators,
a user specifies which buffers are stored in tensor accessible memory
 via Halide scheduling primitives.
\toolname{} relies on Halide's vectorization pass to lower a user's program to a vectorized representation.
The tile extractor annotates the IR program with data movements, generates an EqSat program, and sends the program to egglog.
After egglog finishes optimizing, the tile extractor parses the optimized program back into Halide and performs post-processing.
Finally, the optimized program is passed back to resume Halide's compilation passes, which generate the final machine code.

\toolname{} uses combinations of three existing Halide IR nodes to represent linear tensor operations. All three nodes arise naturally from vectorizing loops in Halide. The first is \verb|Ramp|, which takes a base, a stride, and a length, and represents a vector containing a linear sequence. When a loop is vectorized, instances of the loop variable are replaced with a \verb|Ramp| representing all values of the loop index. The base and stride may themselves be vectors, in which case \verb|Ramp| is the flattened concatenation of a linear sequence of vectors. This occurs when two nested loops are both vectorized. The second IR node is \verb|Broadcast|, which can be thought of as a \verb|Ramp| with a stride of zero. Like \verb|Ramp|, its argument may itself be a vector, where it represents the concatenation of copies of that vector. A broadcast of a value \textit{v} by a factor of \textit{n} is written tersely as \verb|x|\textit{n}\verb|(|\textit{v}\verb|)|. To load and store small tensor tiles, nested combinations of ramps and broadcasts are used as indices to vector load and store instructions. The third relevant IR node is \verb|vector_reduce_add|, which sums fixed-size groups of adjacent scalars to produce a smaller vector. These appear when a reduction is vectorized along the reduction dimension. Some combinations of these IR nodes that represent common tensor operations are shown in Figure~\ref{fig:halide-ir-nodes}.

\begin{figure}
\begin{lstlisting}[style=cpp, mathescape, basicstyle=\ttfamily\footnotesize]
A[ramp(ramp(0, 8, 4), x4(1), 8)]
vector_reduce_add(
    A[x8(ramp(0, 1, 3))]
        * B[ramp(ramp(0, 1, 3), x3(1), 8)],
    3)
vector_reduce_add(
    A[ramp(x6(ramp(0, 1, 3)), x18(3), 4))]
        * B[x4(ramp(ramp(0, 6, 3), x3(1), 6))],
    3)
\end{lstlisting}
\caption{The fully-vectorized representation in Halide IR of a 4x8 transpose, a 3-tap convolution of a signal of size 8, and the multiplication of a 4x3 matrix with a 3x6 matrix,  respectively.}
\label{fig:halide-ir-nodes}
\end{figure}

In the rest of this section, we demonstrate \toolname{} by exploring MatMul in detail. Note that a core goal of \toolname{} is to handle more complex computational patterns that are not literal MatMul (e.g., resampling). However, even straightforward
MatMul is already non-trivial to support in compilers.

\subsection{Scheduling MatMul}

Consider the following MatMul in Halide\footnote{Halide follows the indexing convention of OpenGL, where the innermost dimension is written first (the opposite of C-like languages).}:

\begin{lstlisting}[style=cpp, basicstyle=\ttfamily\footnotesize]
ImageParam A(BFloat(16), 2), B(BFloat(16), 2); 
Var x, y; RDom r(0, 16);
mm(y, x) = 0.f;
mm(y, x) += cast<float>(A(r, x))
            * cast<float>(B(y, r));

mm.in().bound(x, 0, 32).bound(y, 0, 16);
\end{lstlisting}


Line 1-6 describes a MatMul program in Halide,
 and the last line constrains the computation to be
 between a 16x32 matrix and a 32x16 matrix.
The user then writes the following schedule to compile this MatMul program to the AMX accelerator.
The schedule (1) uses the \verb|store_in| directive to ask \verb|mm| to be stored in AMX tile register
 and (2) vectorizes relevant computation:
 
\begin{lstlisting}[style=cpp, basicstyle=\ttfamily\footnotesize]
mm.store_in(MemoryType::AMXTile)
  .compute_at(mm.in(), x);
mm.vectorize(x, 16).vectorize(y, 16);
mm.update().atomic()
  .vectorize(x, 16).vectorize(y, 16)
  .vectorize(r, 32);
mm.in().vectorize(x, 16).vectorize(y, 16);
\end{lstlisting}

This schedule has four statements.
The first statement defines the storage of \verb|mm|
 and creates a wrapper buffer \verb|mm.in()| (in memory) 
 that loads the computed matrix from \verb|mm|.
The rest of the statements \verb|vectorize| the three stages in the computation with corresponding tile sizes: initialization (\verb|mm|), MatMul (\verb|mm.update()|), 
 and loading to the wrapper buffer (\verb|mm.in()|).

\begin{revision}
At this step, the user specifies the target hardware of each computation stage using Halide scheduling directives. This defines the specifications that the instruction selection needs to follow.
For example, it is the instruction selector's job to find available AMX instructions that compute \textit{mm} in AMX tile registers.
In our case, 
This can be mapped to
Intel AMX's \verb|TDPBF16PS| instruction, which does exactly $AB+C$
 for \verb|BFloat16| matrices of this shape.
\end{revision}

\subsection{Matching MatMul}

Guided by the provided schedule, Halide lowers the MatMul program to Halide's intermediate representation using a series of compiler passes including simplification and vectorization.
The resulting IR is shown in \autoref{fig:before-egglog}
 (minus the highlighted data movement nodes).

\toolname{} injects data movement nodes \verb|mem_to_amx| and \verb|amx_to_mem| to indicate 
 program locations where data movements between hosts and accelerators happen.
While Halide's IR does not distinguish between computations in different memories and implicit inject data movement instructions during codegen, \toolname{} needs to explicitly differentiate them:
The E-graph should not treat a MatMul computed in memory  as being equivalent to a MatMul in a tensor register, since using the result of the latter requires first loading the data to memory.

\begin{figure*}
\begin{lstlisting}[style=cpp, emph={mem_to_amx, amx_to_mem},emphstyle={\bfseries\color{DarkPurple}}, basicstyle=\ttfamily\footnotesize]
matmul[ramp(0, 1, 512)] = mem_to_amx( x512(0.f) );
matmul[ramp(0, 1, 512)] = mem_to_amx(
    (float32x512)vector_reduce_add(
        (cast<float32x8192>(A[ramp(x512(0), x512(32), 16) + x256(ramp(0, 1, 32))])
        * x16(cast<float32x256>(B[ramp(ramp(0, 16, 32), x32(1), 16)])))
    ) + matmul[ramp(0, 1, 512)] );
matmul_wrapper[ramp(0, 1, 512)] = amx_to_mem( matmul[ramp(0, 1, 512)] );
\end{lstlisting}
\caption{Halide IR of the example MatMul before equality saturation.
Highlighted in purple are the data movement nodes \toolname{} has injected.
The three statements can be mapped to AMX instructions \texttt{tile\_zero}, \texttt{tile\_matmul}, and \texttt{tile\_load}, respectively.}
\label{fig:before-egglog}
\end{figure*}

\toolname{} prints this program as egglog expressions and sends them to egglog along with a set of predefined egglog rules.
To identify accelerator-leveraging opportunities, egglog looks for the following pattern:

\begin{lstlisting}[language=egglog, basicstyle=\ttfamily\footnotesize]
(= [e] (Add [C] (VectorReduceAdd 512
  (Mul (Cast (Float32 8192) [lhs]) 
       (Cast (Float32 8192) [rhs])))))
(= [lhs] (Load (BFloat16 8192) [A]
  (Ramp (Broadcast (Ramp [A-base] 1 32) 16) 
        (Broadcast [A-stride] 512) 16))))
(= [rhs] (Load (BFloat16 8192) [B]
  (Broadcast (Ramp (Ramp [B-base] [B-stride] 32)
                   (Broadcast 1 32) 16) 16)))
\end{lstlisting}

It looks for a pattern \verb|e| with the form $C+\sum \textit{lhs}\cdot \textit{rhs}$, where \textit{lhs} and \textit{rhs} are expressions that load matrices A and B with certain access patterns.
The access patterns are expressions above with nested \verb|Ramp|s and \verb|Broadcast|s.

The access patterns for \textit{lhs} and \textit{rhs} can be read as follows: both matrices are enumerated with three loop variables (corresponding to the three nesting levels of \verb|Ramp|/\verb|Broadcast|). The outermost loop dimension enumerates over \verb|A| and is constant over \verb|B|, which corresponds to variable $x$ in the source program. The second loop enumerates over \verb|B| and is constant over \verb|A|, so it corresponds to \verb|y|. The innermost loop co-iterates \verb|A| and \verb|B| and corresponds to the reduction domain $r$.
The (pointwise) multiplication between \verb|lhs| and \verb|rhs| creates a vector with 8192 lanes, which is then reduced to 512 lanes via \verb|VectorReduceAdd|.
This is exactly the semantics of a MatMul.

While the pattern we try to match directly follows MatMul,
 Halide lowers the input program to a different shape.
In \autoref{fig:before-egglog}, matrix A is loaded with index \texttt{\allowbreak ramp(broadcast(..), ..) + broadcast(ramp(..), ..)}, 
 while the expected pattern has shape \texttt{\allowbreak ramp(\allowbreak broadcast(\allowbreak ramp(..), ..), ..)}.
Similarly, the B matrix is loaded with \texttt{\allowbreak ramp(ramp(..), ..)} and then broadcasted, again not matching our pattern.
This mismatch is caused by Halide's aggressive simplification passes that run throughout compilation.
While a natural pattern for matrix A would expect three levels of nesting, Halide un-nests it during simplification, resulting in two shallower terms.
For matrix B, Halide's simplifier has converted a load of a broadcast index to broadcasting the loaded value instead, since this is cheaper.
In other words, the local optimization performed by Halide's simplifier obscures the computation pattern, making tensor computation less recognizable. This is an instance of the compiler phase-ordering problem.

Halide's existing support for AMX accelerator implements complex logic to cover each individual post-simplification pattern for programs of interest.
Its pattern matcher is thus obscure and not scalable.

\toolname{} solves this problem by introducing another set of axiomatic rules.
For instance, to push \texttt{broadcast} inside \texttt{load},
it defines the following egglog rule:
\begin{lstlisting}[language=egglog, basicstyle=\ttfamily\footnotesize]
(rewrite 
  (Broadcast (Load [type] [name] [index]) [lanes]) 
  (Load (MultiplyLanes [type] [lanes]) [name]
        (Broadcast [index] [lanes])))
\end{lstlisting}
\texttt{MultiplyLanes} computes the correct type for the new load, and it will be rewritten to a concrete type via supporting rules like the following:
\begin{lstlisting}[language=egglog, basicstyle=\ttfamily\footnotesize]
(rewrite (MultiplyLanes (Float32 [l]) [x]) 
         (Float32 (* [l] [x]))
\end{lstlisting}
EqSat automatically finds the right order to apply these rules so that our MatMul pattern gets matched.

Similarly, \toolname{} has a set of axioms between nested \texttt{Ramp}/\texttt{Broadcast} expressions and their unnests. 
Thanks to EqSat, we can express target applications using principled patterns and,  independent of application patterns,  define a set of axiomatic rules, without worrying about optimizations interfering with one another.

\subsection{Lowering MatMul to Hardware}

So far we have successfully recognized the MatMul pattern, but there is a caveat.
AMX expects matrix B to be in the VNNI format, 
 a specific data layout that groups B's rows by 2 and interleaves elements from rows in a group.
Indeed, to use AMX for our MatMul example requires data shuffling:
\begin{lstlisting}[style=cpp, basicstyle=\ttfamily\footnotesize]
B_vnni(x%2, y, x/2) = B(y, x);
\end{lstlisting}
 
Halide's existing AMX support assumes the data is already stored in the VNNI format,
so it actually cannot lower our MatMul example.
In fact, the AMX applications it supports must have the following form:
\begin{lstlisting}[style=cpp, basicstyle=\ttfamily\footnotesize]
mm(y, x) += cast<float>(A(r, x))
        * cast<float>(B_vnni(r%2, y, r/2));
\end{lstlisting}
Yet there are many applications that do not directly fit this pattern.
As a result, application developers need to
 rewrite their computation to account for accelerator-specific data layouts and swizzling, 
 failing Halide's promises of decoupling algorithms and schedules.
 
\toolname{} solves this problem with a set of application-specific rules.
Application-specific rules encode domain knowledge about MatMul-mappable optimization patterns.
To support different layouts for matrix B,
 \toolname{} uses a binary predicate relation \texttt{amx-B-tile}.
\begin{lstlisting}[language=egglog]
(relation amx-B-tile (Expr Expr))
\end{lstlisting}
The first argument to \texttt{amx-B-tile} is an AMX-mappable matrix, 
 and the second argument is 
 an expression that loads this matrix to AMX tile registers (with potential shuffling).
The full rule for discovering matrices in the standard layout as AMX-mappable is shown in the Appendix (first rule, Figure 10b~\cite{hardboiled-arxiv}).
In its query, this rule looks for matrix B's loading pattern in the standard layout.
For each matched loading pattern, 
 the rule runs its action that (1) constructs an expression that shuffles B matrix,
 stores the result in a temporary buffer,
 and loads the buffer to memory, and (2) declares a fact of the predicate relation \verb|amx-B-tile| that 
 the constructed matrix
 holds the same data as the original matrix but in the desired layout.

Finally, \toolname{} uses lowering rules to emit AMX instructions.
The full rule is shown in the Appendix (first rule, Figure 10a~\cite{hardboiled-arxiv}).
It looks for expressions of the form $C+\sum A\times B$, queries relations \verb|amx-A-tile| and \verb|amx-B-tile| for matrices $A$ and $B$ in the expected layout, and constructs an equivalent MatMul expression that uses AMX.

\subsection{Discussion}

So far, we have shown an end-to-end walkthrough of \toolname{}'s flexible support of tensor operation patterns via EqSat. A more detailed description of the intermediate representation and rules is presented in the Appendix~\cite{hardboiled-arxiv}. In this section, we discuss several practical considerations in the implementation of \toolname{}.

\subsubsection{Supporting Nvidia Tensor Cores}

Besides Intel AMX, \toolname{} also supports Warp-level Matrix Multiply-Accumulate (WMMA) instructions of Nvidia Tensor Cores.
Different from CPUs,
Nvidia's CUDA programming model organizes computations into GPU blocks and threads, 
 and a WMMA instruction executes synchronously across a warp of 32 threads.
To support WMMA instructions, 
 the tile extractor wraps each WMMA-related statement with a warp-level parallel loop \texttt{for\_gpu\_lanes (thread\_id\_x, 0, 32) \{ \}}. 
\toolname{} relies on Halide's \texttt{FuseGPUThreadLoop} pass to fuse multiple warp-level loops into a single loop.

Under WMMA's semantics each thread holds a fragment of the tile, while Halide by default stores allocations outside the warp-level execution, so it is shared across threads in a warp.
Therefore, \toolname{} also pushes the WMMA buffer allocation inside the warp-level execution and scales down the allocation size to be per thread.


\subsubsection{Rule Schedule}

A \emph{rule schedule} in egglog controls when and how often rules run.
Currently, \toolname{} uses a simple rule schedule, where it runs a fixed number of iterations of axiomatic, app.-specific, and lowering rules, and between each iteration, runs supporting rules to fixpoints. This is because supporting rules like type analysis always saturate in finite steps.
%

\subsubsection{Cost Model}

After equality saturation, egglog needs to extract the optimal program from the E-graph, which requires a cost model. \begin{revision}Unlike previous work on auto-scheduling~\citep{ansor,tensorir_2022}, which use sophisticated cost models, our cost model is simple, AST-size based. This is sufficient because, in our setting, the schedule is explicitly provided by the developer (e.g., with primitive \verb|store_in(AMXTile)|). This makes instruction selection a hit-or-miss problem.\end{revision}

\subsubsection{Supporting More Flexible Patterns in \toolname{}}

\begin{revision}
Thanks to the rule-based design, it is easy to extend \toolname{} with support for new lowering patterns, and with EqSat, we do not have to worry about the interaction of rules leading to the phase-ordering problem.

\begin{shepherd}
\toolname{} has implemented two categories of rules that lower to hardware:
MatMul-like rules, which support both the standard and VNNI layouts, and convolution-like rules, which use a generalized Toeplitz transformation (\autoref{sec:convolution}) to lower convolution-like patterns, including single-channel convolution, upsampling and downsampling, to MatMul hardware intrinsics.
We found these two categories of rule sufficient to cover a range of unconventional workloads, like those considered in this paper.
\end{shepherd}
\end{revision}

\subsubsection{Implementing Tensor Accelerator Kernels Using \toolname{}}

\begin{revision}
\begin{shepherd}
With \toolname{}'s flexible pattern support, users often do not need to contort their Halide programs (the algorithms) into MatMul patterns that are in direct correspondence to the hardware's capability; instead, they can express the algorithms in their natural forms.
Users then update schedules to be accelerator-efficient and use the scheduling primitive \verb|store_in| to determine which part of the computation to run on tensor accelerators.

All \toolname{} schedules in the end-to-end case studies, except resampling (\autoref{sec:resampling}), are implemented in this way. Expressing resampling as linear matrix operations requires global transformations, while, as an instruction selector, \toolname{} only emits local, vector-lane-level shuffling instructions. Interestingly,  re-expressing resampling in a form more amenable to tensor accelerators also provides speedups on non-tensor hardware by increasing arithmetic intensity.
\end{shepherd}
\end{revision}


\section{Evaluation on ML Workloads}

Before diving into applications that do not typically use tensor accelerators, we first evaluate \toolname{} on traditional ML workloads as a sanity check of our system. In particular, we focus robustness and performance.

For performance measurements on Tensor Cores, we use an \begin{shepherd}AWS EC2 p4de.24xlarge instance with Nvidia A100 80GB SXM\end{shepherd}\footnote{\shepherd{We use a data center GPU here as it is representative of large-scale ML workloads. In \autoref{sec:case-study}, we evaluate our image processing case studies using a desktop with an RTX 4070 SUPER, since image processing workloads are typically run on consumer hardware.}}. Since we did not have access to a CPU that supports Intel’s AMX, we used the Intel Software Development Emulator (version 9.38) to validate the correctness of \toolname{}-generated code, focusing on functional correctness rather than performance.

\paragraph*{Robustness}



\begin{figure*}[t]
    \centering
    \begin{minipage}[t]{0.32\textwidth}
        \centering
        \captionsetup{
            type=table,
            belowskip=-100pt
        }
        \captionof{table}{Support for MatMul schedules from Intel's Optimization Reference Manual in two layouts.}
        \label{tab:intel-amx-new}
        \resizebox{\textwidth}{!}{
        \begin{tabular}{|>{\centering\arraybackslash}p{3.2cm}
                        >{\centering\arraybackslash}p{1.5cm}
                        >{\centering\arraybackslash}p{1.8cm}|}
        \hline
        \textbf{Implementation} & \textbf{VNNI} & \textbf{Standard} \\
        \hline
        Reference impl.        & \checkmark & \checkmark \\
        Loop reordering        & \checkmark & \checkmark \\
        Preloading matrix A    & \checkmark & \checkmark \\
        Preloading matrix B    & \checkmark & $\times$ \\
        Software pipelining    & $\times$   & $\times$ \\
        \hline
        \end{tabular}
        }
    \end{minipage}
    \hfill
    \begin{minipage}[t]{0.32\textwidth}
        \centering
        \includegraphics[width=\textwidth]{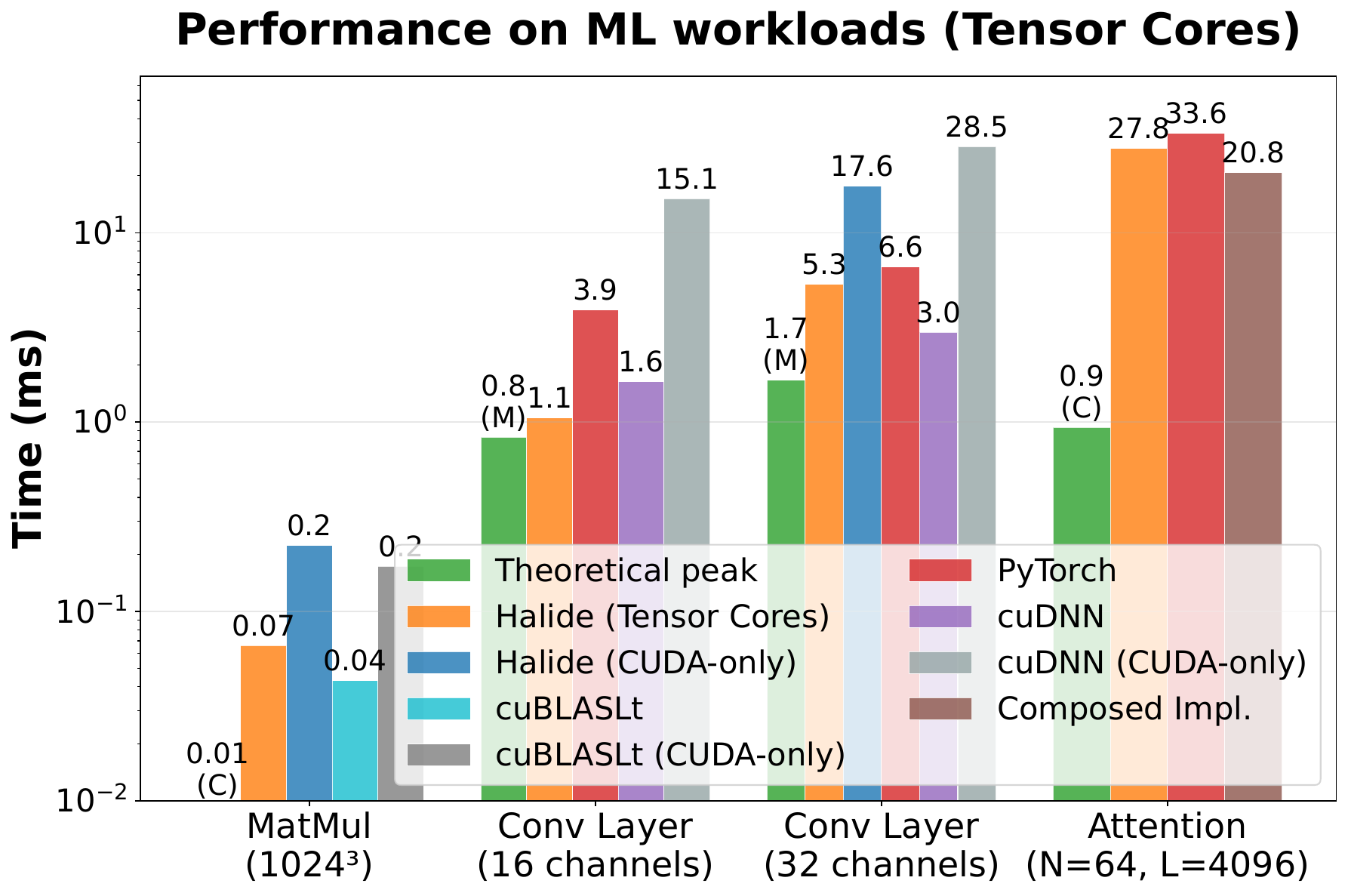}
        \captionof{figure}{Performance comparison on ML workloads.}
        \label{fig:ml-workloads}
    \end{minipage}
    \hfill
    \begin{minipage}[t]{0.32\textwidth}
        \centering
        \includegraphics[width=\textwidth]{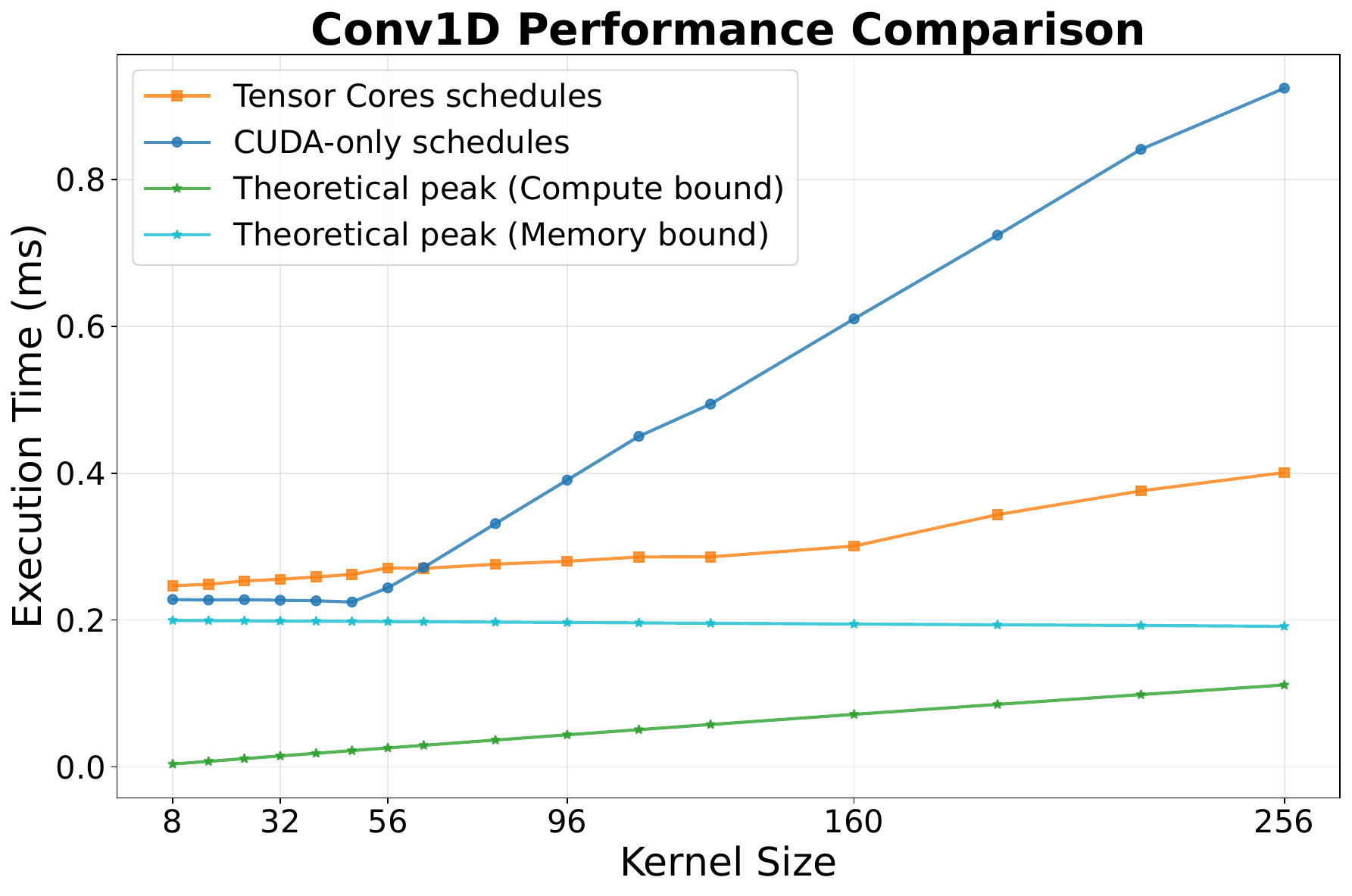}
        \captionof{figure}{Performance comparison on 1D convolution.}
        \label{fig:conv1d}
    \end{minipage}
    
    \vspace{1em}

    \begin{minipage}[t]{0.32\textwidth}
        \centering
        \includegraphics[width=\textwidth]{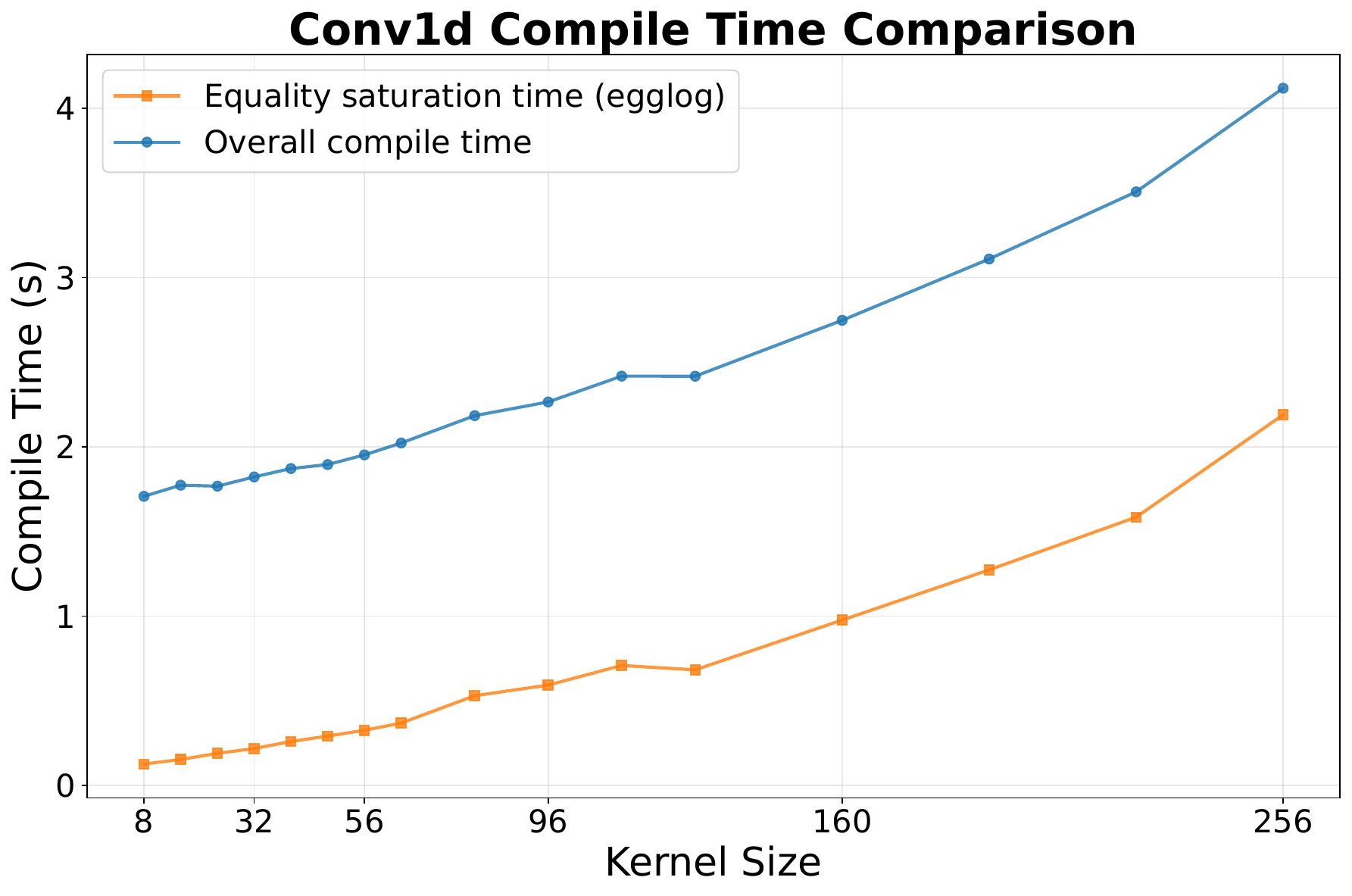}
        \captionof{figure}{Kernel compilation time of 1D convolution.}
        \label{fig:compile-time}
    \end{minipage}
    \hfill
    \begin{minipage}[t]{0.32\textwidth}
        \centering
        \includegraphics[width=\textwidth]{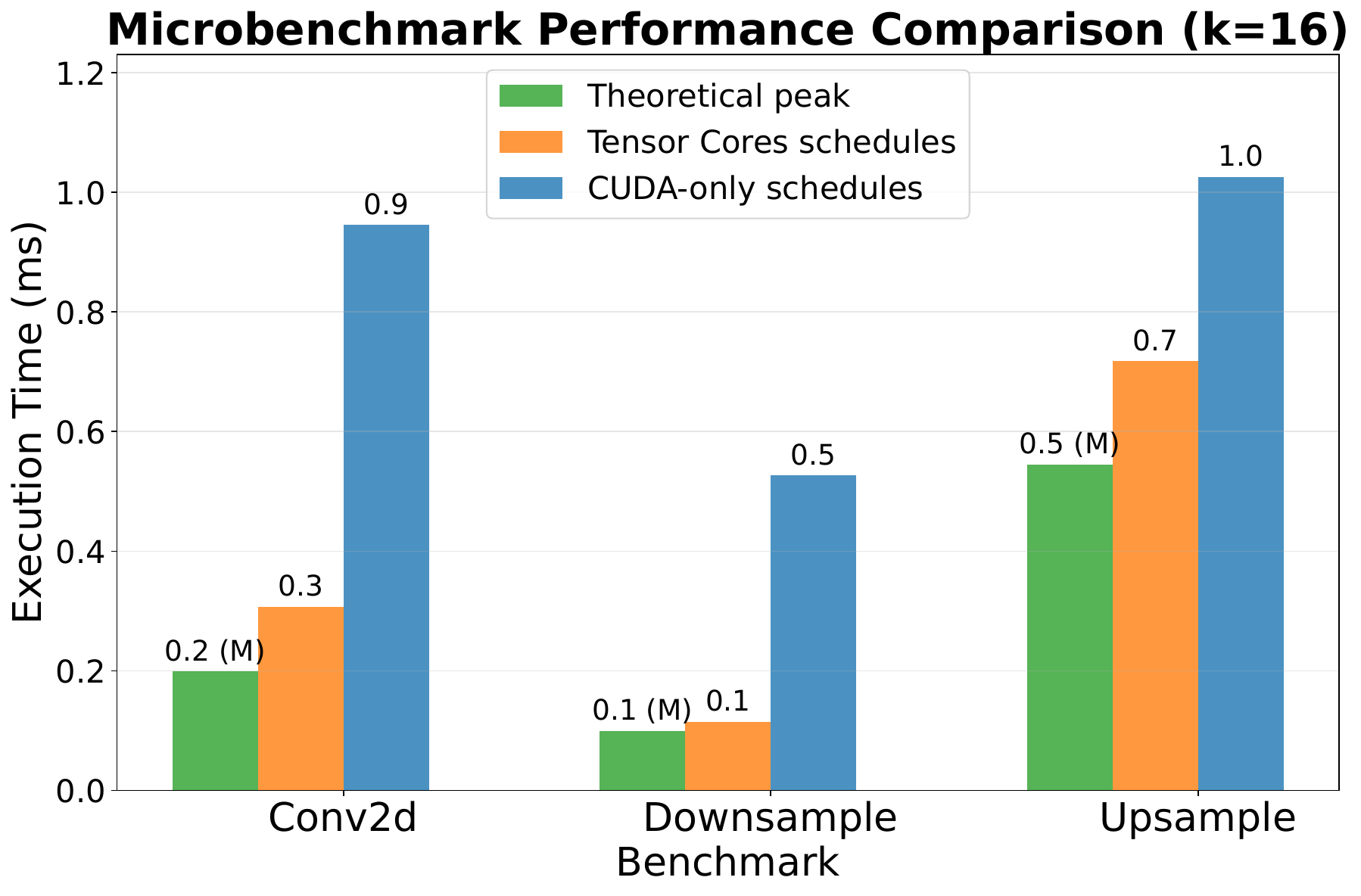}
        \captionof{figure}{Performance comparison on microbenchmarks with kernel size 16.}
    \label{fig:2d-k16}
    \end{minipage}
    \hfill
    \begin{minipage}[t]{0.32\textwidth}
        \centering
        \includegraphics[width=\textwidth]{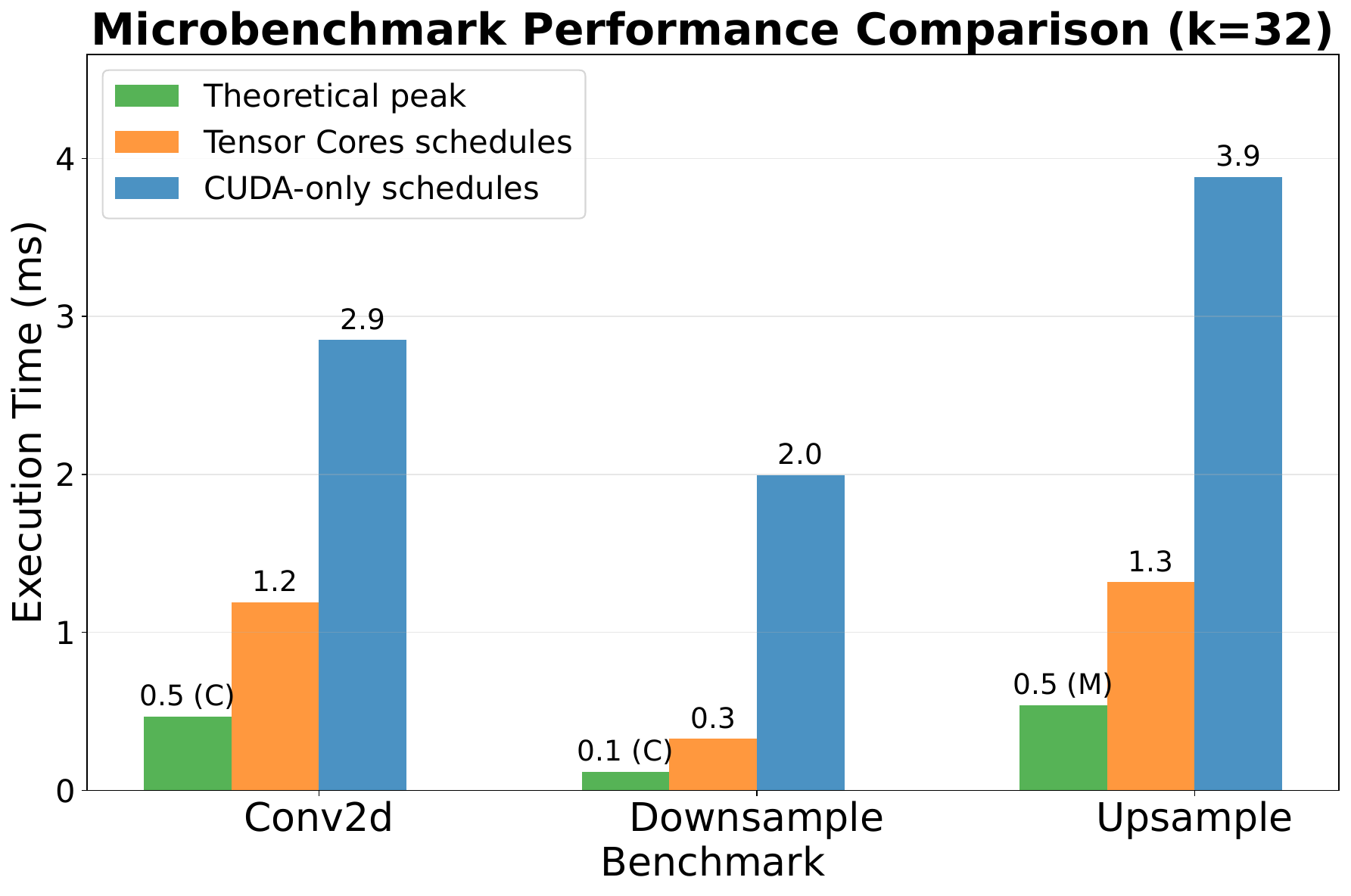}
        \captionof{figure}{Performance comparison on microbenchmarks with kernel size 32.}
    \label{fig:2d-k32}
    \end{minipage}
\end{figure*}

We use a collection of MatMul implementations from Intel’s Optimization Reference Manual~\citep[Chapter 20.5.5]{intel_optimization_manual_2024} and 
reimplement them as \toolname{} schedules. 
\autoref{tab:intel-amx-new} summarizes each schedule and its compatibility with \toolname{}.
The reference manual assumes inputs are stored in the VNNI format expected by AMX instructions and that elements are pre-swizzled. In addition to evaluating this layout, we also test the more conventional ``standard'' layout, where operands are stored in row-major order. In this setting, \toolname{} discovers when swizzling is required and injects the necessary swizzling code without user annotations or schedule changes. 

Across the VNNI-format schedules, \toolname{} supports all cases listed in the manual except software pipelining. Software pipelining requires fine-grained interleaving of load/store and compute instructions that cannot be expressed in Halide’s scheduling model. Under the standard layout, \toolname{} supports all from VNNI except the one that preloads matrix~B.
In this case, it is ambiguous whether the preloaded data should be swizzled.

\paragraph*{Performance}

\begin{revision}
To validate that our system produces high-quality code for standard workloads, we benchmark three canonical operators: matrix multiplication, convolutional layer, and scaled dot-product attention. For each workload we compare against strong Tensor Core baselines and Halide implementations that target CUDA without Tensor Cores. Our goal here is not to exhaustively tune Halide schedules or to outperform vendor libraries, but rather to demonstrate that \toolname{} can work with provided Halide schedules to yield competitive performance on well-understood problems. 
The results are summarized in \autoref{fig:ml-workloads}.

\begin{shepherd}
For reference, we have also added the optimal runtime for each benchmark based on A100 Tensor Core's theoretical TFLOPs (156 TFMAs for FP16) and bandwidth (2 TB/s)~\citep{nvidia2020ampere}. Each bar is labeled with either (C) or (M) to indicate whether it is compute-bound or memory-bound.
\end{shepherd}

We evaluated GEMM on 1024$\times$1024 matrices. Our baselines are \texttt{cuBLASLtMatmul}, NVIDIA’s optimized GEMM routine (v12.6.4), and a CUDA-only schedule from Halide's repository\footnote{The official Halide repository has a set of manually-tuned example applications. The Halide GEMM implementation is available at \url{https://github.com/halide/Halide/tree/main/apps/cuda_mat_mul}.}. Our Tensor Core-enabled schedule reduced runtime from 223 us to 66 us, a 3.36$\times$ speedup. Our generated code is about 1.5$\times$ slower compared to \texttt{cuBLASLtMatmul}.


We next benchmarked a standard convolutional layer with bias and activation, matching the forward-propagation primitive commonly used in neural networks. 
We compare against two vendor baselines: \texttt{torch::nn::Conv2d} (PyTorch v2.5.1), and \texttt{cudnn\-Convolution\-Bias\-Activation\-Forward} (Nvidia’s cuDNN v9.1.0), as well as a CUDA-only Halide schedule drawn from Halide's repository\footnote{The implementation is available at \url{https://github.com/halide/Halide/tree/main/apps/conv_layer} and only supports 32 channels and beyond.}. On input size 4096$\times$64$\times$64$\times$16, our Tensor Core schedule achieved 1.05 ms, outperforming PyTorch by 3.7$\times$, and 1.6$\times$ faster than cuDNN. On input size 4096×64×64×32, our schedule ran in 5.34 ms, making it 3$\times$ faster than the CUDA-only baseline, 1.2$\times$ faster than PyTorch and 1.8$\times$ slower than cuDNN.

Finally, we benchmarked scaled dot-product attention, which consists of three stages: (1) computing $QK^{\mathsf{T}}$ and scaling, (2) applying a softmax across keys, and (3) multiplying the resulting probabilities with $V$. We considered two vendor baselines: a PyTorch implementation using \texttt{torch.nn.functional} primitives, and a composed version combining cuBLAS for MatMuls, cuDNN softmax for normalization, and a lightweight custom CUDA kernel for scaling. We do not compare against highly optimized kernels such as FlashAttention, since they incorporate algorithmic and data layout optimizations that are orthogonal to Tensor Core uses; our Halide implementation follows the na\"ive attention algorithm for an apples-to-apples comparison, focusing on proper Tensor Core utilization.

On input size $N=64$, $L=4096$, $D=64$, our Tensor Core implementation achieved 27.8 ms, outperforming the PyTorch baseline by 1.2$\times$, while trailing the composed cuBLAS+cuDNN baseline by about 1.3$\times$. 

These results demonstrate that our system can produce hardware-accelerated kernels that are competitive with vendor baselines. We emphasize that our goal is not to beat cuBLAS or cuDNN
but to show that our support for tensor accelerators is effective. 
\end{revision}



\section{Case Study}
\label{sec:case-study}




Using Halide equipped with \toolname{} for tensor instruction selection,  we are able to map many classical operations in signal and image processing to tensor accelerators. In this section, we present a series of case studies, focusing on mapping to the Nvidia Tensor Cores in a commodity desktop GPU, the GeForce RTX 4070 SUPER.
We study the following applications: \circled{1} 1D/2D convolution, \circled{2} upsampling by 2, \circled{3} downsampling by 2, \circled{4} resampling by a non-integer factor, \circled{5} recursive filtering, and \circled{6} DCT-based denoising. Source code for all case studies is available online\footnote{\url{https://github.com/yihozhang/cgo2026-hardboiled-artifact}}.

Among them, the first three applications are micro benchmarks that can be reformulated as linear matrix operations via local transformations on the kernel matrix, so \toolname{} emits shuffling instructions to transform it into Tensor Cores MatMul instructions. 

The last three applications are end-to-end applications that can benefit from the use of tensor accelerators for small matrix multiplications or convolutions in key places within larger pipelines. In the interest of space we will describe these at a higher level. Readers are encouraged to see the source code for the differences between the CUDA-only schedules and the tensor core schedules.

In addition to proving the generality and usefulness of \toolname{}, our main research question in these case studies is whether tensor accelerators can provide meaningful speedups for these kinds of applications, so our primary baselines will be our best-effort optimized Halide CUDA schedules that do not use Tensor Cores.

\def\rx{{r_x}}
\def\ry{{r_y}}
\def\dx{{d_x}}
\def\dy{{d_y}}

\subsection{Convolution}
\label{sec:convolution}

We first consider 1D convolution of the form $O(x) = \sum_{0\leq \rx< l} I(x+\rx)\cdot K(\rx)$ and 2D convolution of the form $O(x, y) = \sum_{0\leq \rx < l, 0\leq \ry < m} I(x+\rx,y+\ry)\cdot K(\rx,\ry)$, where the kernel $K$ has size $l$ and $m\times l$ respectively.
Unlike multi-channel convolutions used in convolutional layers, they cannot be mapped to MatMul by e.g., im2col~\citep{im2col}, which would degenerate to a matrix-vector multiplication.

Let us first review how to reformulate 1D convolutions into linear operations. Consider values of $O$ at positions $(x, y),\ldots, (x, y+k)$. It can be computed as

\begin{center}
\scalebox{0.8}{
$\displaystyle
\begin{bmatrix}
    O(x), \ldots , O(x+k-1)
\end{bmatrix} =  
\begin{bmatrix}
    I(x), \ldots,  I(x+k+l-1)
\end{bmatrix}\times A_K
$
}
\end{center}

where $A_K$ is a $k\times (k+l)$ Toeplitz matrix defined by 

\begin{center}
\scalebox{0.8}{
$\displaystyle
A_K(x, y) = \begin{cases} 
&K(y-x) \quad\text{ if } 0\leq y-x < l\\
&0\quad\text{ otherwise } \end{cases}
$
}
\end{center}

To turn this vector-matrix multiply into a matrix-matrix multiply, we can load $h$ (overlapped) vectors of length $k+l$ from $I$:
\begin{align}
\label{eqn:conv}
\scriptstyle
\begin{bsmallmatrix}
    O(x) & \ldots& O(x+k-1) \\
    O(x+k)& \ldots & O(x+2k-1) \\
     &\vdots&  \\
    O(x+(h-1)k)
    &\ldots& O(x+hk-1)
\end{bsmallmatrix} &=  
\begin{bsmallmatrix}
    I(x) 
    &
    \ldots 
    & 
    I(x+k+l-1)
    \\
    I(x+2k) 
    & 
    \ldots 
    & 
    I(x+2k+l-1)
    \\
    &
    \vdots
    & 
    \\
    I(x+(h-1)k)
    &
     \ldots
     & 
     I(x+hk+l-1)
\end{bsmallmatrix}\\
& \qquad \qquad \qquad \times A_K
\end{align}

\begin{shepherd}
\toolname{} automatically constructs the Toeplitz matrix for suitable patterns, so 1D convolution can be naturally expressed as the following Halide algorithm:
\end{shepherd}

\begin{lstlisting}[style=cpp, basicstyle=\ttfamily\footnotesize]
ImageParam K(Float(16), 1), I(Float(16), 1);
RDom rx;
conv(x) = 0.f;
conv(x) += cast<float>(K(rx))
           * cast<float>(I(x + rx));   
output(x) = conv(x);
\end{lstlisting}
Both the kernel and the input are in \verb|float16|, and the output is in \verb|float32|, a layout that WMMA instructions on Tensor Cores support.

To direct \toolname{} to lower the convolution to Tensor Cores, the user writes a schedule that divides the computation into $256$-pixel segments and a reduction axis of $8$, and requests each segment to be computed on WMMA accumulator registers.
In code, this looks as follows:

\begin{lstlisting}[style=cpp, basicstyle=\ttfamily\footnotesize]
output.split(x, x, xi, 256)
      .vectorize(xi);
conv.compute_at(output, x)
    .store_in(WMMAAccumulator)
    .split(x, x, xi, 256)
    .vectorize(xi);
conv.update()
    .split(x, x, xi, 256).split(rx, rx, rxi, 8)
    .reorder(rxi, xi, rx, x)
    .atomic().vectorize(xi).vectorize(rxi);
\end{lstlisting}
The computation of each segment is equivalent to multiplying a $16\times 32 $ input matrix consisting of 32 (overlapped) vectors with an $8\times 16$ kernel matrix, producing an $32\times 8$ tile as output. This maps to WMMA's $32\times16\times 8$ MatMul. \begin{shepherd}More details on how \toolname{} maps this 1D convolution pattern to hardware can be found in Appendix B~\cite{hardboiled-arxiv}.\end{shepherd}

To tensorize a 2D convolution, we can parametrize over one reduction axis, which reduces the problem to 1D convolutions: 
define function $O_\ry(x,y)$ parametrized by $\ry$ as $\sum_{0\leq \rx< l} I(x + \rx,y+\ry)\cdot K(\rx,\ry)$ where $\ry$ and $y+\ry$ are effectively constants, and $O(x,y) = \sum_{0\leq \ry <m} O_\ry(x, y)$.
This parametrization step, when reflected in Halide schedules, is equivalent to leaving $\ry$ as a serial outer loop.

These schedules produce inner loops that perform two matrix loads, one matrix multiply, and one matrix store. To achieve a higher arithmetic intensity, we can additionally unroll along any axis on which one of the operands depends but the other does not. For convolution on an image, we unroll over multiple rows of the output to amortize the cost of loading the convolution kernel, which does not vary along that axis.

\autoref{fig:conv1d} shows the result for 1D convolution, comparing a manually tuned Halide CUDA schedule. Starting at around $k=64$, the CUDA-only schedule transitioned from bandwidth-limited to compute-limited, while the Tensor Core schedule remained bandwidth-limited, culminating in a speedup of 2.3$\times$ at $k=256$. For 2D convolution, the Tensor Core schedule achieves a 3.1$\times$ speedup at $k=16$ (\autoref{fig:2d-k16}) and 2.4$\times$ speedup at $k=32$ (\autoref{fig:2d-k32}).

\begin{shepherd}
The figures also show the theoretical performance limits based on RTX 4070 SUPER's theoretical TFLOPs and bandwidth\footnote{
\begin{shepherd}
For RTX 4070 SUPER, we use a peak theoretical Tensor Core compute of 36 TFMAs, obtained by scaling the numbers advertised for RTX 4090 \cite{nvidia2022ada} proportionately to the reduced number of Tensor Cores, since they have the same Tensor Core specification. The peak bandwidth of RTX 4070 SUPER is advertised to be 504.2 GB/s.\end{shepherd}
}.
This is computed using the algorithmic description of the benchmark and is oblivious to the redundant and unnecessary computations introduced by the use of accelerators (such as Toeplitz transformations)%
\footnote{\begin{shepherd}For example, the theoretically optimal amount of compute for 1D convolution is calculated as $(4096-k)\cdot 4096\cdot k$ for kernel size $k$, and the optimal IO is calculated as the sum of input size and output size.
\end{shepherd}}, to avoid being dependent on the specific schedules and transformations used.
As a result, even schedules that use Tensor Cores perfectly cannot reach the performance limit shown in the figures.
\end{shepherd}

We also measured run time spent in equality saturation for 1D convolution.
Since we unroll along the reduction dimension, larger kernel sizes mean longer programs.
Unlike many EqSat applications that suffer from blowups, \toolname{} exhibits a manageable run time because (1) individual tensorized statements are usually small and (2) \toolname{}'s search for tensor-leveraging opportunities is guided by user-provided schedules and requires a smaller search space than traditional super-optimizers.

\subsection{Downsampling and Upsampling by an Integer Factor}

We consider the task of downsampling (strided convolution) by a factor of 2: 
\begin{center}
\scalebox{0.8}{
$\displaystyle \sum_{\rx,\ry}I(2x+\rx,2y+\ry)\cdot K(\rx,\ry).
$}
\end{center}

By parameterizing over $\ry$ via loop reordering, we are left with downsampling along one axis: $\sum_{\rx}I(2x+\rx,y_1)\cdot K(\rx, y_2)$.
This requires constructing the following Toeplitz-like coefficient matrix of size $k\times (2k+l)$:

\begin{center}
\scalebox{0.8}{
$\displaystyle A^\textit{down}_K(x,y)=
\begin{cases} 
&K(y-2x) \quad\text{ if } 0\leq y-2x < l\\
&0 \end{cases}$
}
\end{center}

On the other hand, upsampling by a factor of 2 has the following form:

\begin{center}
\scalebox{0.8}{
$\displaystyle O(x,y)=\sum_{\rx,\ry}I\left(\left\lfloor \frac{x}{2}\right\rfloor+\rx,\left\lfloor \frac{y}{2}\right\rfloor+\ry\right)\cdot K(2\rx+x\%2,2\ry+y\% 2)
$
}
\end{center}

We can simplify the indexing by separating it into phases and treating this as a multiphase filter:

\scalebox{0.8}{
\begin{minipage}{\linewidth}
\begin{align*}
K^{\textit{phase}}(x, y,\dx, \dy) &= K(2x+\dx, 2y+\dy)\\
O^{\textit{phase}}(x, y,\dx, \dy) &= \sum_{\rx, \ry}I(x + \rx, y+\ry) \cdot K^{\textit{phase}}(\rx, \ry,\dx, \dy) \\
O(x, y) &= O^{\textit{phase}}\left(\left\lfloor \frac{x}{2}\right\rfloor, \left\lfloor \frac{y}{2}\right\rfloor, x\% 2, y \% 2\right)
\end{align*}
\end{minipage}
}


Note that $O^\textit{phase}$ is already in the form of a convolution that \toolname{} supports, but this would be memory inefficient since the output is stored in an interleaved form in $O^\textit{phase}$. To be more cache-friendly, we store $O^\textit{phase}$ compactly by reordering the storage of $\dx,\dy$ as the innermost dimensions:
\begin{lstlisting}[style=cpp]
O_phase.reorder_storage(dx, dy, x, y);
\end{lstlisting}

An $h\times k\times 2\times 2$ tile of $O^\textit{phase}$ with a reduction axis of length $l$ can be computed as

\scalebox{0.8}{
\begin{minipage}{\linewidth}
\begin{align*}
\begin{bmatrix}
O^\textit{phase}(x, y, 0, 0) &\ldots &O^\textit{phase}(x+k-1, y, 1, 1)\\
O^\textit{phase}(x+k, y, 0, 0) &\ldots &O^\textit{phase}(x+2k-1, y, 1, 1)\\
&\vdots&\\
O^\textit{phase}(x+(h-1)k, y, 0, 0) &\ldots &O^\textit{phase}(x+hk-1, y, 1, 1)\\
\end{bmatrix}\\
=\begin{bmatrix}
    I(x,y )
    &
    \ldots 
    & 
    I(x+\lfloor \frac{k}{4} \rfloor +l-1,y)
    \\
    I(x+2k,y) 
    & 
    \ldots 
    & 
    I(x+2\lfloor \frac{k}{4}\rfloor + l-1,y)
    \\
    &
    \vdots
    & 
    \\
    I(x+(h-1)k,y)
    &
     \ldots
     & 
     I(x+h\lfloor \frac{k}{4} \rfloor + l+l-1,y)
\end{bmatrix}
\times A^\textit{up}_{K, y}
\end{align*}
\end{minipage}
}

where $A^\textit{up}_{K, w}$ is a $\lfloor \frac{k}{4}\rfloor\times \left(\lfloor \frac{k}{4} \rfloor + l\right)$ matrix:

\begin{center}
\scalebox{0.8}{
$\displaystyle
A^\textit{up}_{K, w}(x, y)= 
\begin{cases}
K^\textit{phase}(y-\left\lfloor \frac{x}{4}\right\rfloor, w, x \% 2, \left\lfloor \frac{x}{2}\right\rfloor\% 2) \\
\quad\qquad \qquad\qquad \text{ if } 0\leq y-\left\lfloor \frac{x}{4}\right\rfloor < l\\
0\quad\text{ otherwise }
\end{cases}
$
}
\end{center}

\begin{shepherd}
\toolname{} recognizes both downsampling and upsampling patterns, where it would generate shuffling instructions that construct the matrix $A^\textit{down}_{K}$ and $A^\textit{up}_{K, w}$.
\end{shepherd}

As in convolution, we added a small amount of additional unrolling to increase the number of matrix multiply instructions per matrix load and store instruction. 
This results in our downsample and upsample implementations, which achieve 4.6$\times$ and 1.4$\times$ speedups respectively at $k=16$, and 6.1$\times$ and 2.9$\times$ speedups at $k=32$.

\subsection{Resampling by a Non-Integer Factor}
\label{sec:resampling}
In some settings, it is necessary to quickly resample an image down to a smaller fixed size, regardless of its original size. This happens when generating thumbnails for a gallery, or when preparing inputs to an ML model trained at a fixed size. Unlike the integer-factor resizing above, there is no single filter bank that is used per output pixel. Mathematically, resampling is done by convolving the high-resolution samples with a continuous pre-filter to make a continuous signal, then sampling that continuous signal at a lower rate. The pre-filter is sized according to the output sampling rate and designed to reject any frequencies that cannot be represented at reduced resolution, so its footprint on the input can be quite large. A popular choice for a high-quality pre-filter is a three-lobed Lanczos~\cite{lanczos3}, defined as $sinc(x)sinc(x/3)$ for $x \in [-3, 3]$, where $sinc(x) = sin(\pi x)/(\pi x)$.

Evaluating this Lanczos filter repeatedly is expensive. Fortunately, resizing can be separated into vertical followed by horizontal resizing. When resizing an image vertically, each column undergoes the same linear transformation. This linear transformation can be precomputed as a single sparse matrix applied to all columns, amortizing the cost of evaluating the Lanczos kernel across columns. The same can be done for horizontal resizing. This sparse matrix is rectangular, with a non-zero band running down the diagonal with a known bounded width. An easy representation for this sparse matrix is the index of the first non-zero followed by the non-zero entries, and this is the one used by the \texttt{resize} application in Halide's repository\footnote{\url{https://github.com/halide/Halide/tree/main/apps/resize}}. 

However, we can also treat this matrix as block-sparse: we can give entire groups of rows the same starting index, and expand the width of the band as necessary. This introduces some unnecessary multiplications by zero. Still, it is an overall win even without tensor accelerators because it lets us tile the computation in a memory-efficient way, reducing the number of repeated loads of the same input values. We found aligning groups of 16 rows in this way sped up the original \texttt{resize} application by about $3\times$.

\begin{table}[t]
  \centering
  \vspace{0.5em}
  \caption{Runtimes for resizing a 2048$\times$2048 image down by a non-integer factor using a three-lobed Lanczos pre-filter. The geomean speedup is 1.47$\times$.}
  \label{table:resize_results}
  \begin{tabular}{|ccc|}
    \hline
    \textbf{Output size} & \textbf{CUDA-only (us)} & \textbf{Tensor core (us)} \\
    \hline
    143$\times$143 & 111 & 79 \\
    245$\times$245 & 110 & 73 \\
    450$\times$450 & 113 & 74 \\
    921$\times$921 & 145 & 102 \\
    \hline
  \end{tabular}
\end{table}

We modified the \texttt{resize} application to be block-sparse in this way, further optimized its CUDA schedule, and benchmarked resampling a four-megapixel RGB image by several non-integer factors. Results are shown in Table~\ref{table:resize_results} under the column ``CUDA-only''. These kernels achieve high utilization of both compute and memory bandwidth, both ranging between 60\% and 90\% depending on the size and resampling direction.

We then changed the schedule to perform the small matrix multiplication stages on Tensor Cores. Adding compute using Tensor Cores makes the kernels bandwidth-limited. This schedule achieves only 10\% utilization of the Tensor Cores, but it is enough to yield a geometric mean overall speedup of $1.47\times$ across the output sizes tested.

This application demonstrates that low-arithmetic-intensity workloads such as resizing images can still benefit from Tensor Cores, despite using them at a fraction of their peak throughput.

\subsection{Recursive Filtering}

In audio processing, recursive filters are used as a core primitive for a variety of effects. A simplified second-order recursive filter on an input signal $X$ can be expressed as:

\begin{center}
\scalebox{0.8}{
$\displaystyle Y_t = X_t + \alpha Y_{t-1} + \beta Y_{t-2}$
}
\end{center}

Each output is a linear combination of the next input and the previous two outputs. A direct implementation is necessarily entirely serial. Audio signals may have multiple channels that can be processed in parallel, but for stereo audio this number is merely two. Fortunately, a variety of techniques exist for transforming recursive filters into forms that admit parallel computation. First, they can be computed in parallel tiles using the method of Hoppe et al.~\cite{gpu-rec-filter}. This technique tiles the domain and computes the original filter on each tile in parallel. As additional passes, it then fixes up each tile using the contribution from previous tiles. This provides a parallel outer loop. 

We can also achieve a parallel inner loop using a technique called Scattered-lookahead interpolation (SLA) by Parhi et al.~\cite{scattered-lookahead}. Given a dilation factor $d$, it gives a recipe for decomposing the recursive filter into a non-recursive convolution of size $2d + 1$ followed by a dilated recursive filter of the form:

\begin{center}
\scalebox{0.8}{
$\displaystyle Y_t = X_t + \alpha' Y_{t-d} + \beta' Y_{t-2d}$
}
\end{center}

In this form, the recursive step can operate in parallel over groups of $d$ outputs. We found that a combination of these two techniques was most effective for implementing a recursive filter on GPU, with inter-block parallelism provided by the first technique, and intra-block parallelism provided by the second. Transformed in this way, our optimized Halide CUDA implementation takes 67.5~us to process $2^{21}$ stereo samples (about 50 seconds of audio). The runtime is roughly two thirds in the recursive step and one third in the fix-up step required by Hoppe et al. The convolution is fused into the recursive step. The fix-up step is limited by global memory bandwidth, achieving 90\% of peak, while the recursive step is limited by bandwidth from L1, achieving 92\% of peak. We found performance was maximized at $d = 8$ and tiles of size $1024$.

Without changing the algorithm, we changed the Halide schedule to use Tensor Cores for the convolution: we vectorize across the reduction in the convolution and across groups of 256 outputs, and place the results in shared memory to be immediately used by the recursive step. \toolname{} can recognize the convolution in this form, similar to \autoref{sec:convolution}, and constructs a Toeplitz matrix for our convolution kernel ahead of time. This schedule reduced overall runtime to 58~us, with all the savings coming from the recursive step (the fix-up kernel is unchanged). As the recursive kernel was L1-bandwidth-limited, the speedup is not due to harnessing additional compute, but rather because the Tensor Core schedule is easier on the memory subsystem. The Tensor Cores are at a mere 8\% utilization.

This application demonstrates how ad hoc usage of Tensor Cores inside non-ML pipelines can provide meaningful speedups, despite only using a small fraction of their peak capabilities.

\subsection{DCT-Based Denoising}

A traditional approach to reduce noise in images is transform-domain coring. An image is divided into overlapping tiles, with a soft windowing function applied. Each tile then undergoes some linear transformation designed to concentrate the signal into a few bins while leaving the noise spread across many bins. Any bin with a small magnitude is then set to zero. The transform is then inverted, and overlapping tiles are blended into a single output. In our case, we will use a discrete cosine transform (DCT), in line with the most notable use of transform-domain coring---the JPEG file format~\cite{jpeg}. We will use both a direct DCT, and the fast DCT described by Plonka et al.~\cite{plonka}. The tile size is set to be 16$\times$16. 

For the direct variant, the DCT is performed by treating each tile as a 16$\times$16 matrix, and multiplying it by the DCT matrix to transform down each column. The result is transposed, and multiplied by the DCT matrix again to transform along each row. After coring, the same is done in reverse along rows and columns using the inverse DCT matrix, resulting in four MatMuls per tile. These are all fused into a single kernel. The fast DCT variant has the same structure, but the MatMuls are replaced by a fully unrolled 16-point fast DCT run along the rows or columns. After the transform kernel, a second blending kernel is launched to blend overlapping output tiles and undo the windowing.

Our optimized Halide CUDA implementation takes 277~us and 76~us to denoise a one megapixel three-channel image using the direct and fast approaches respectively. Focusing on the fast variant, time is split roughly equally between the transform kernel and the blending kernel. The transform kernel is balanced, achieving 75\% of peak compute and 77\% of peak memory bandwidth, while the blending kernel is entirely bandwidth-limited.

We rescheduled the direct variant to use Tensor Cores for the MatMuls. This yields a total runtime of 68~us.
By using Tensor Cores, the brute-force DCT is four times faster than using CUDA cores alone, and is about 10\% faster than the fast DCT despite doing $3.6\times$ more floating-point operations. The transform kernel is bandwidth-limited, achieving 85\% of peak bandwidth from DRAM and 88\% of peak bandwidth from L1. Tensor Core utilization is 28\%.
 
This transform kernel fuses four matrix multiplications with a non-linear operation (the coring) in between, yet it is still bandwidth-limited. If implemented using separate calls to a library to perform the matrix multiplication steps, this fusion would be lost, and the runtime would scale up according to the number of kernel launches required. Being able to ad hoc schedule individual pipeline stages onto Tensor Cores in a fused kernel is more efficient than having to break out of Halide and call a library routine as a separate kernel launch.

\section{Related Work}
\label{sec:relatedwork}

\subsection{Program Optimization Using Equality Saturation}


Several works have been proposed on targeting domain-specific accelerators with EqSat. For example, Diospyros~\cite{diospyros} and JOKer~\cite{joker} use EqSat-based rewrite rules to select vector instructions for digital signal processors (DSPs).
The most relevant to us is Glenside~\cite{glenside,3la}, which uses low-level rewrite rules to map tensor programs to DL accelerators.
Both Glenside and \toolname{} work at the instruction selection level, 
but
Glenside is a standalone project, while \toolname{} works with Halide. 
The integration with Halide allows users to develop hardware-accelerated kernels in a high-level scheduling DSL.

Similar to \toolname{}, which considers how to retrofit EqSat to existing compilers, DialEgg\cite{dialegg} integrates egglog with MLIR in a dialect-agnostic way. Caviar~\cite{caviar} boosts the optimizing rewrite rules in Halide with EqSat and shows it can prove pre-conditions that Halide cannot.
\begin{revision}
LIAR~\citep{liar} rewrites program patterns in a minimalistic DSL into hardware-accelerated instructions using EqSat.
\end{revision}

\subsection{Compilers for Hardware Accelerators}

HalideTCU \cite{halide_tcu_2020} was an independent attempt to add Tensor Cores support to Halide. It added a high-level \verb|tensor_core| method to the IR that just passes parameters to WMMA instructions, 
but does not support scheduling.
Many other frameworks support Tensor Cores, especially within the ML programming community.  
TVM \cite{tvm} exposes Tensor Core intrinsics but requires the developer to manually allocate memory in the correct format and write adapter methods to target the intrinsics directly \cite{tvm-wmma-tutorial}. 
TileLang \cite{tilelang} extends TVM with high-level abstractions for MatMul and other common ML operations, but it still requires the developer to allocate memory in the correct format and to manually copy data around. 
Triton \cite{triton} is a Python DSL for developing GPU kernels and provides a library of primitives that map to hand-written CUDA code. When the library does not match their needs, developers have to write their own intrinsic (e.g., to fuse data movement and computation). 
The RISE language for rewrite-based performance tuning \cite{rise_2022} was recently extended to target Tensor Cores.

\begin{revision}
Recent work like Ansor and TensorIR has focused on auto-scheduling for \usl{}s. TensorIR \cite{tensorir_2022} is especially relevant to us, since it targets Tensor Cores and applies a similar tile abstraction to TVM's IR.
\toolname{} is different from this work in several aspects: first, auto-scheduling aims to find good schedules that automatically map high-level workloads onto tensor cores. In contrast, we want to enable tensor core usage from human-written schedules, which means \toolname{} must correctly map code onto tensor cores in precisely the way it is told to by a human, rather than making a best effort at finding \emph{any} good mapping to tensor instructions. The additional constraints make the problem easier in some respects and harder in others. 
Moreover, work like TensorIR focuses on traditional ML tasks such as GEMM and convolution layers, while we focus on workloads that may leverage tensor accelerators but have not done so (potentially due to a lack of suitable tooling). 
\end{revision}

\toolname{} can be viewed as solving the lifting problem from lower-level tensor operations to high-level primitives like MatMul.
Tensorize~\cite{tensorize_2025} uses symbolic traces to lift algorithms in legacy codebases to high-level tensor programs.
Rake \cite{rake} improves instruction selection for DSPs by first lifting Halide IR to a higher-level Uber-Instruction IR, and Pitchfork~\citep{pitchfork} takes a similar lift-then-lower approach but uses offline rewrite rule synthesis.

\section{Conclusion}
\begin{revision}
In this work, we add support for tensor accelerators to the Halide scheduling language in the \toolname{} system.
\toolname{} gives Halide users fine-grained direct control over how portable algorithm code is mapped to tensor accelerators in a way that composes cleanly with the existing scheduling language. 
It builds on Halide's existing vector IR, and uses equality saturation to select tensor instructions from Halide expressions that have been vectorized along multiple axes at once.
We have shown that \toolname{} enables ad hoc use of tensor accelerators in signal processing applications that have not traditionally used them, and that this provides meaningful speedups.

\end{revision}

\section*{Acknowledgment}
\addcontentsline{toc}{section}{Acknowledgment}

We would like to thank the anonymous reviewers and shepherd for their valuable feedback and to friends who share their insights during this work, including Shoaib Kamil, Alex Reinking, Zachary Tatlock, and Zihao Ye.
This material is based upon work supported by the National Science Foundation
under Grant No.~2232339 and 2312195.

\section*{Data Availability}

Additional data related to this
publication may be found in the repository at \url{https://doi.org/10.5281/zenodo.17810573}~\cite{ahmad_2025_17810573}.

\bibliographystyle{plain}

\def\UrlBreaks{\do\/\do-}

\bibliography{references}

\newpage

\appendix

\subsection{Implemenetation of \toolname{}}

\label{sec:ir}

\begin{figure}
    \centering
\small    
\begin{align*}
\text{Stmt } s &::= \texttt{store}(\textit{str}, e_{\textit{idx}}, e_{\textit{val}})
  \mid  \texttt{eval}(e) \\
\text{Expr } e &::= \texttt{load}(\textit{str}, t, e_{\textit{idx}}) 
  \mid \texttt{cast}(t, e)  
  \mid \texttt{call}(f, \vec{e})
  \\
  &\quad 
  \mid \textit{op}(\vec{e})
  \mid \texttt{ramp}(e_\textit{ini}, e_\textit{inc}, n) 
  \mid \texttt{broadcast}(e, n) \\
  &\quad \mid \texttt{vector\_reduce\_add}(n, e) \\
  &\quad \mid {\texttt{loc\_to\_loc}(l_1, l_2, e)}
  \mid x 
  \mid v \\
\text{Location } l &::= \texttt{Mem} 
  \mid \texttt{AMX} 
  \mid \texttt{WMMA} \\
\text{Type } t &::= \texttt{bfloat16}^n 
  \mid \texttt{float32}^n 
  \mid \cdots \\
\text{Arith.~Op. } \textit{op} &::= + \mid - \mid * \mid \ /_{\ } \mid \cdots \\
\text{Intrinsics } \textit{f} &::= \texttt{tile\_load} \mid \texttt{tile\_matmul} \mid \cdots \\
\text{Variable } x &::= a \mid b \mid \cdots \mid {\texttt{ExprVar}(e)} \\
\text{Value } v &\in [\![ \texttt{bfloat16}]\!] \cup [\![ \texttt{float32}]\!]\cup \cdots
\end{align*}
    \caption{Syntax of \toolname{}'s internal representation}
    \label{fig:ir}
\end{figure}
Our IR is shown in \autoref{fig:ir}. A program is a sequence of statements.
A \texttt{store} statement takes a buffer name, an index pattern, and a vector expression and stores the vector in the buffer according to the index.
An \texttt{evaluate} statement evaluates a given expression and is useful for intrinsics with side effects, e.g., \texttt{tile\_store}.

A \texttt{load} operator takes a buffer name, the result type, and an index pattern and does a vectorized load similar to \texttt{store}.
A \texttt{cast} operator converts a value to a compatible type. 
A \texttt{call} operator describes a (tensor) intrinsic call,
 and arithmetic operators $\textit{op}$ are interpreted using standard arithmetic over numbers and pointwise over vectors.

Both \texttt{broadcast} and \texttt{ramp} expressions are primitives for building indexing vectors.
Semantically, $\texttt{ramp}(e_\textit{ini}, e_\textit{inc}, n)$ concatenates vectors 
$e_\textit{ini}, e_\textit{ini}+e_\textit{inc},\ldots,e_\textit{ini}+(n-1)e_\textit{inc}$,
and $\texttt{broadcast}(e, n)$ represents a concatenation of $n$ copies of $e$.
A $\texttt{vector\_reduce\_add}$ operator\footnote{Halide supports vector reduction of any associative-commutative operators. For this paper we only need $+$ as the aggregation operator.} takes an aggregation factor $n$ and a vector expression $e$ with $k$ lanes.
It requires $k$ to be divisible by $n$ and partitions vector $e$ into $k$ subvectors, and sums within each subvector to derive an aggregated vector with $k$ lanes.
\texttt{loc\_to\_loc} marks data movement between memories and accelerator registers.

Besides standard variables, \toolname{} also uses expression variables to represent temporary buffers.
\texttt{ExprVar} takes a vector expression and denotes a pointer to a buffer that stores the result of evaluating that expression.
\toolname{} uses \texttt{ExprVar} to store swizzled matrices.

\section{Optimization rules}

\label{sec:rules}

\begin{figure*}[h!]
\begin{minipage}[t]{0.5\textwidth}
\begin{subfigure}[t]{\textwidth}
\begin{lstlisting}[language=egglog, basicstyle=\small\ttfamily]
(rule ( ;; lowering matmul
  (= [e] (Add [C] (VectorReduceAdd 512
    (Mul (Cast (Float32 8192) [A])
         (Cast (Float32 8192) [B])))))
  (amx-A-tile [A] [amx-A])
  (amx-B-tile [B] [amx-B])
) (
  (let [new-e] (Call "tile_matmul" 
    (vec-of (Mem2AMX [C]) [amx-A] [amx-B])))
  (union [e] (AMX2Mem [new-e]))
))
(rule ( ;; lowering tile store
  (= [s] (Store [buffer] (AMX2Mem [tile]) [index]))
  (= [index] (Ramp (Ramp [base] 1 16) 
                 (Broadcast [stride] 16) 
                 16))
) (
  (let [new-s] (Evaluate 
    (Call "tile_store" (vec-of 
      (Var [buffer]) [base] [stride] [tile]))))
  (union [s] [new-s])
))
\end{lstlisting}
\caption{Lowering rules}
\label{fig:lowering-rules}
\end{subfigure}

\vspace{1em}

\begin{subfigure}[t]{\textwidth}
\begin{lstlisting}[language=egglog, basicstyle=\small\ttfamily]
(rewrite (Broadcast (Broadcast [x] [l1]) [l2])
         (Broadcast [x] (* [l1] [l2])))
(rewrite (Broadcast [x] 1) [x])
(rewrite (Broadcast (Load [type] [name] [index]) 
                    [lanes]) 
         (Load (MultiplyLanes [type] [lanes])
               [name] 
               (Broadcast [index] [lanes])))
(rewrite (Broadcast (Cast [type] [expr]) [lanes]) 
         (Cast (MultiplyLanes [type] [lanes]) 
               (Broadcast [expr] [lanes])))
(rewrite (Add (Ramp [base] [stride] [r-lanes])
              (Broadcast [x] [b-lanes]))
         (Ramp 
           (Add [base] 
             (Broadcast [x] (/ [b-lanes] [r-lanes]))) 
           [stride] 
           [r-lanes])
    :when ((= (% [b-lanes] [r-lanes]) 0)))
(rewrite (Ramp [x] [s] 1) [x])
\end{lstlisting}
\setcounter{subfigure}{2}
\caption{Axiomatic rules}
\label{fig:axiomatic-rules}
\end{subfigure}
\end{minipage}
\hfill
\begin{minipage}[t]{0.5\textwidth}
\begin{subfigure}[t]{\textwidth}
\begin{lstlisting}[language=egglog, basicstyle=\small\ttfamily]
(rule ( ;; loading B in the standard layout
  (= [orig-B] (Load (BFloat16 8192) [B-name] [B-index]))
  (= [B-index] (Broadcast
    (Ramp (Ramp [B-base] [B-stride] 32)
          (Broadcast 1 32)
          16)
    16))
) (
  (let [type-B] (BFloat16 256))
  (let [load-B] (Load [type-B] [B]
    (Ramp (Ramp [B-base] 1 32) 
          (Broadcast [B-stride] 16) 
          16)))
  (let shuffle-B (ExprVar (Call "KWayInterleave" 
    (vec-of 2 32 [load-B])))
  (let amx-B (Call "tile_load"
    (vec-of [shuffle-B] 0 64)))
  (amx-B-tile [orig-B] [amx-B])
))
(rule ( ;; loading B in the VNNI layout
  (= [orig-B] (Load (BFloat16 8192) [B] [B-index]))
  (= [B-index] (Broadcast
    (Ramp (Ramp (Ramp [B-base] 1 2) 
                (Broadcast [B-stride] 2)
                16)
          (Broadcast 2 32)
          16)
    16))
) (
  (let [amx-B] (Call "tile_load"
    (vec-of (Var [B]) [B-base] [B-stride])))
  (amx-B-tile [orig-B] [amx-B])
))
\end{lstlisting}
\setcounter{subfigure}{1}
\caption{Application-specific rules}
\label{fig:application-specific-rules}
\end{subfigure}

\vspace{1em}

\begin{subfigure}[t]{\textwidth}
\begin{lstlisting}[language=egglog, basicstyle=\small\ttfamily]
(function MultiplyLanes (Type i64) Type)
(rewrite (MultiplyLanes (Float32 [l]) [x]) 
         (Float32 (* [l] [x])))
(rewrite (MultiplyLanes (BFloat16 [l]) [x]) 
         (BFloat16 (* [l] [x])))
(relation has-type (Expr Type))
(rule ((= e (FloatImm v))) 
      ((has-type e (Float32 1))))
(rule ((= e (Ramp [inner] [s] [l])) 
       (has-type [inner] [ty]))
      ((has-type [e] (MultiplyLanes [ty] [l]))))
(rule ((= e (Broadcast [inner] [l])) 
       (has-type [inner] [ty])) 
      ((has-type [e] (MultiplyLanes [ty] [l]))))
\end{lstlisting}
\setcounter{subfigure}{3}
\caption{Supporting rules}
\label{fig:supporting-rules}
\end{subfigure}
\end{minipage}
\caption{Example egglog rules in \toolname{} by category.}
\end{figure*}

The heavy lifting of optimization is performed via rewrite rules in egglog.
We classify rules into four categories: lowering rules, application-specific rules, axiomatic rules, and supporting rules.

\subsubsection{Lowering rules}
\label{sec:lowering}

Lowering rules lower high-level tensor patterns to accelerator intrinsics.
\autoref{fig:lowering-rules} presents the lowering rules for
 MatMul and storing the results to memory.
The lowering rule captures the MatMul pattern $AB+C$ while 
 abstracting away the representation details with \texttt{amx-*-tile} relations.
\toolname{} uses additional application-specific rules to populate \texttt{amx-*-tile}
 relations (\autoref{sec:app-rules}).

The lowering rule for AMX MatMul does not assume the result is
 already marked as being stored in AMX registers. 
Instead, it unions the original expression with \lstinline[language=egglog]{(AMX2Mem[ new-e])},
 so even if the user does not indicate the result should be stored in AMX registers (e.g., intermediate computations that users do not get a handle to schedule),
 \toolname{} may still leverage AMX when possible and load the result to memory.
In the case the user has required the result to be stored in memory, 
 \toolname{} cancels out \texttt{AMX2Mem} with rule \lstinline[language=egglog]{(rewrite (Mem2AMX} \lstinline[language=egglog]{(AMX2Mem   [e])) [e])}.

The second rule in \autoref{fig:lowering-rules} eliminates \texttt{AMX2Mem} by 
looking for \texttt{store}s from AMX to memory and union them with
 a call to the \texttt{tile\_store} intrinsics.

\subsubsection{Application-specific rules}
\label{sec:app-rules}

Application-specific rules extend and complement lowering rules 
 with diverse accelerator-leveraging MatMul patterns.
\autoref{fig:application-specific-rules} shows two rules supporting different B matrix patterns.
The first rule searches for the access pattern of the B matrix in the standard, dense layout.
For each such B, it uses the \texttt{KWayInterleave} intrinsics to shuffle the B matrix,
 materializes the result in memory, and \verb|tile_load| is to AMX tile register.
The second rule searches for the loading pattern of the VNNI layout verbatim.
The outer \texttt{broadcast} corresponds to the \verb|i| dimension of \verb|A(k, i) * B(k%2, j, k/2)|
 (where the A matrix is enumerated while the index over B stays the same), 
 and the three inner \texttt{ramp}s correspond to \verb|k/2|, \verb|j|, \verb|k| respectively.

Both rules use the \texttt{amx-B-tile} relation that abstracts over the application-specific details,
 which gets queried by the lowering rules for AMX.
This allows us to avoid rule duplication by implementing only one rule for lowering to hardware.
\toolname{} also has a number of rules that combine the application-specific patterns and hardware-specific lowering.
\autoref{sec:case-study} will showcase a rule that converts an image convolution into MatMul 
 by constructing a Toeplitz matrix.

\subsubsection{Axiomatic rules}
\label{sec:axiomatic}

\toolname{} uses axiomatic rules to make pattern matching robust to syntactic variations.
Some vector-related axiomatic rules are shown in \autoref{fig:axiomatic-rules}.
For instance, the first rule reads that two consecutive \texttt{broadcast}s can be flattened into one.
\toolname{} also implements additional arithmetic axioms such as commutativity.
We do not implement associativity as it can make the E-graph grow exponentially
 and our use case does not yet require associativity.

One consideration in writing these rules is to decide what equivalent expressions to populate,
 similar to the quantifier instantiation problem in SMT solving~\citep{quantifier-instantiation}.
For instance, the inverse of the \texttt{broadcast}-flattening rule above, which creates nested broadcasts,
 is also a valid rule, and it is helpful in several settings.
But the inverse rule is not immediately applicable: we cannot infer what \texttt{(* l1 l2)} is before we know what \texttt{l1} and \texttt{l2} is!
\toolname{} uses two heuristics. First, we use sibling terms as hints. For example, \toolname{} uses the following rule\footnote{In our implementation, we parametrize binary operators as arguments to a more general \texttt{Bop} constructor, so the presented rule would work for any binary operator, instead of just addition.}:
\begin{lstlisting}[language=egglog]
(rewrite 
  (Add (Ramp [x] [s] [l1]) (Broadcast [a] [l2]))
  (Add (Broadcast (Broadcast [a] (/ [l2] [l1])) [l1]) 
       (Ramp [x] [s] [l1]))
  :when ((> [l2] [l1])
         (= 0 (% [l2] [l1]))))
\end{lstlisting}

This rule says whenever a \texttt{broadcast} expression has a sibling \texttt{ramp} expression
 with a different argument lanes\footnote{Note that for the expression to be well-defined, the two expressions must have the same number of lanes, but \texttt{l1} may not be equal to \texttt{l2} because \texttt{a} and \texttt{x} may have different lanes.},
 we can nest the \texttt{broadcast} expression so that the \texttt{broadcast} expression has the same number of lanes as the \texttt{ramp}.
In standard term rewriting, this is a pointless rule, because it complicates the original expression. However, with equality saturation, this rule can help other rules to make further simplifications.
For example, this rule enables the following rule, which pushes the addition inside the base of a \texttt{ramp}.
\begin{lstlisting}[language=egglog]
(rewrite (Add (Ramp [x] [s] [l]) (Broadcast [a] [l]))
         (Ramp (Add [x] [a]) [s] [l]))
\end{lstlisting}

Our second heuristic is to capture common cases. 
For instance, because of the VNNI layout, if the B matrix has size 2xM,
 the buffer degenerates to a 1D vector of length $2M$,
 and the access pattern becomes a single \texttt{ramp}.
In \toolname{}, we introduce the following two rules to recover the general pattern from the degenerate case\footnote{Because {\color{DarkPurple}\texttt{x}} on the LHS is unbounded, it needs to be grounded by a supporting relation {\texttt{IsExpr}}.}:
\begin{lstlisting}[language=egglog]
(rewrite [x] (Broadcast [x] 1) :when ((IsExpr [x])))
(rewrite (Ramp [e] 1 [l]) 
  (Ramp (Ramp [e] 1 2) (Broadcast 2 2) (/ [l] 2)) 
\end{lstlisting}

\subsubsection{Supporting rules}
\label{sec:supporting}

Many rules we have seen so far depend on certain helper relations, such as \texttt{MultiplyBy}, which multiplies the lanes of a vector type by a given number. 
Here we describe two other helper analyses used in \toolname{}. These analyses are expressed as Datalog-style deductive rules in egglog.

\emph{Type checking.}\quad We implement type checking as a few deductive rules over e-classes. The type information is used in several rules.

\emph{Shape tracking.}\quad AMX and WMMA accept tensors of different dimensions, and different tensor dimensions require different intrinsics for loading/storing. 
We track tensor dimensions using relations \verb|AMXShape| and \verb|WMMAShape|, which are populated when MatMul intrinsics are emitted and queried when synthesizing intrinsics for loading tensors from/to memory.

\section{Tile extractor pass}
\label{sec:tile-extractor}
\toolname{} relies on the tile extractor pass  to pre- and post-process the program and communicate with Halide. 

The tile extractor pass annotates the input program with data movement between memory and accelerator registers. It then looks for allocations marked by the user as being stored in accelerator memory and annotates data movements of these allocations.

Next, the tile extractor collects \verb|store| statements of the input program and prints each \verb|store| statement as egglog expressions. The tile extractor then generates an egglog program with rules described in \autoref{sec:rules} and spawns an egglog process to run the program. 
After EqSat finishes, egglog extracts the optimal program with regard to a cost model and outputs it.
\toolname{}'s current cost model prioritizes terms with smaller sizes. 
The tile extractor parses egglog's output back to the internal representation in C++
and replaces the original \verb|store| statements with optimized programs egglog produces.

The tile extractor lowers potential \verb|ExprVar| nodes in the optimized program by emitting allocations with appropriate sizes and their initializations. Such allocation declarations are lifted as far as possible to maximize their reuses across loop iterations.
Finally, the tile extractor desugars the shuffling intrinsics emitted during EqSat into shuffle operators in Halide IR (omitted in \autoref{sec:ir}).

\subsection{Supporting convolution-like patterns using Toeplitz transformation}
\label{sec:convolution-walkthrough}

Assume the user writes the following program:

\begin{lstlisting}[style=cpp, basicstyle=\ttfamily\footnotesize]
ImageParam K(Float(16), 1),
           I(Float(16), 1);
RDom rx;
conv(x) = 0.f;
conv(x) += cast<float>(K(rx)) * cast<float>(I(x + rx));   
output(x) = conv(x);

output.split(x, x, xi, 256)
      .vectorize(xi)
      .gpu_blocks(x);
conv.compute_at(output, x)
    .store_in(WMMAAccumulator)
    .split(x, x, xi, 256)
    .vectorize(xi);
conv.update()
    .split(x, x, xi, 256)
    .split(rx, rx, rxi, 8)
    .reorder(rxi, xi, rx, x)
    .atomic()
    .vectorize(xi)
    .vectorize(rxi);
\end{lstlisting}

The following IR program shows the convolution expression that Halide lowers to:

\begin{lstlisting}[style=cpp, mathescape, basicstyle=\ttfamily\footnotesize]
conv[ramp(0, 1, 256)] =
  (float32x256)vector_reduce_add(
    float32x2048(I[
      ramp(ramp(rx, 1, 8), x8(1), 256)
    ]) *
    x256(float32x8(K[ramp(rx, 1, 8)]))
  ) + conv[ramp(0, 1, 256)]
\end{lstlisting}

Under the hood, \toolname{} looks for $e$ of the following form that describes a 1D convolution.

\begin{lstlisting}[language=egglog, basicstyle=\ttfamily\footnotesize]
(= [e] (Add [C] (VectorReduceAdd 256
  (Mul (Cast (Float32 2048) 
           (Load (Float16 2048) [I] [index-I])) 
       (Cast (Float32 2048) 
           (Load (Float16 2048) [K] [index-K]))))))
(= [index-I] (Ramp (Ramp [base-I] 1 8)
                 (Broadcast 1 8)
                 256))
(= [index-K] (Broadcast (Ramp [base-K] 1 8) 256))
\end{lstlisting}
and produces a temporary buffer that stores the Toeplitz matrix obtained by \verb|ConvolutionShuffle|-ing kernel $K$.
\begin{lstlisting}[language=egglog, basicstyle=\ttfamily\footnotesize]
(let [new-K] (ExprVar (Call "ConvolutionShuffle"
  (vec-of (Var [K]) [base-K] 16 8))))
\end{lstlisting}
\toolname{} then asserts that the convolution between $I$ and $K$ is equivalent to a MatMul between $I$ loaded using pattern in Equation~\ref{eqn:conv} and \texttt{new-K}.

\begin{revision}
\verb|ConvolutionShuffle| is an intrinsic that we add to Halide to support convolution-like computations.
It constructs the matrix $A_K$ from kernel $K$ and is implemented using LLVM's \verb|shuffle| instructions (as Halide compiles to LLVM).
\end{revision}

After \toolname{}'s pass, the following code is generated.

\begin{lstlisting}[style=cpp, mathescape, basicstyle=\ttfamily\footnotesize]
allocate tempbuf[float16 * 128]
tempbuf[ramp(0, 1, 128)] = 
  ConvolutionShuffle(K, rx, 16, 8)
conv[ramp(0, 1, 8)] = 
  wmma.mma.sync.aligned.row.row.m32n8k16.f32.f32(
    wmma.load.a.sync.aligned.row.m32n8k16.f16(
      I, rx, 8), 
    wmma.load.b.sync.aligned.row.m32n8k16.f16(
      tempbuf, 0, 8), 
    conv[ramp(0, 1, 8)])
\end{lstlisting}

\end{document}